\documentclass[fleqn,usenatbib]{mnras}

\usepackage{newtxtext,newtxmath}
\usepackage{lscape}
\usepackage{multirow}
\usepackage{longtable}

\usepackage[T1]{fontenc}

\DeclareRobustCommand{\VAN}[3]{#2}
\let\VANthebibliography\thebibliography
\def\thebibliography{\DeclareRobustCommand{\VAN}[3]{##3}\VANthebibliography}

\usepackage{graphicx}	
\usepackage{amsmath}	

\title[Planetary systems in the light of asteroseismology]{
Planetary systems in the light of asteroseismology: metallicity threshold for the planetary systems and age-metallicity relation}

\author[M. Y{\i}ld{\i}z, A. Demirkol, S. \"Ortel, T. \c{C}ak{\i}r Alsa\c{c} and T. Daylan]{M. Y{\i}ld{\i}z$^{}$\thanks{E-mail:
mutlu.yildiz@ege.edu.tr}$^1$, A. Demirkol$^2$, S. \"Ortel$^2$, T. \c{C}ak{\i}r Alsa\c{c}$^2$ and T. Daylan$^3$\\
$^1$Department of Astronomy and Space Sciences, Faculty of Science, Ege University, 35100 \.Izmir, Turkey.\\
$^2$Department of Astronomy and Space Sciences, Graduate School of Natural and Applied Sciences, Ege University, 35100 \.Izmir, Turkey.\\
$^3$Department of Physics, Washington University, St. Louis, MO 63130, USA}
\date{Accepted XXX. Received YYY; in original form ZZZ}

\pubyear{2024}

\begin{document}
\label{firstpage}
\pagerange{\pageref{firstpage}--\pageref{lastpage}}
\def\braket#1{\left<#1\right>}
\newcommand{\yildiz}{Y\i ld\i z }
\newcommand{\etal}{et al. }
\newcommand{\wrt}{with respect to }
\newcommand{\logg}{\log(g) }
\newcommand{\numino}{\mbox{\ifmmode{\overline{\nu_{\rm min}}}\else$\overline{\nu_{\rm min}}$\fi}}
\newcommand{\numin}{\mbox{\ifmmode{\nu_{\rm min}}\else$\nu_{\rm min}$\fi}}
\newcommand{\teff}{\mbox{\ifmmode{T_{\rm eff}}\else$T_{\rm eff}$\fi}}
\newcommand{\teffsun}{\mbox{\ifmmode{{\rm T}_{\rm eff{\sun}}}\else${\rm T}_{\rm eff{\sun}}$\fi}}
\newcommand{\numax}{\mbox{$\nu_{\rm max}$}}
\newcommand{\nuH}{\mbox{\ifmmode{\nu_{\rm minH}}\else$\nu_{\rm minH}$\fi}}
\newcommand{\nuL}{\mbox{\ifmmode{\nu_{\rm minL}}\else$\nu_{\rm minL}$\fi}}
\newcommand{\Dnu}{\mbox{$\Delta \nu$}}
\newcommand{\Dpi}{\mbox{$\Delta \upi$}}
\newcommand{\muHz}{\mbox{$\mu$Hz}}
\newcommand{\kepler}{\mbox{{\it Kepler}}}
\newcommand{\corot}{\mbox{{\it CoRoT}}}
\newcommand{\tess}{\mbox{{\it TESS}}}
\newcommand{\gaia}{\mbox{{\it Gaia}}}
\newcommand{\numaxS}{\mbox{$\nu_{\rm max {\sun}}$}}
\newcommand{\MS}{{\rm M}\ifmmode_{\sun}\else$_{\sun}$~\fi}
\newcommand{\MJ}{{\rm M}\ifmmode_{\rm J}\else$_{\rm J}$~\fi}
\newcommand{\RJ}{{\rm R}\ifmmode_{\rm J}\else$_{\rm J}$~\fi}
\newcommand{\Rp}{{\rm R}\ifmmode_{\rm p}\else$_{\rm p}$~\fi}
\newcommand{\Mp}{{\rm M}\ifmmode_{\rm p}\else$_{\rm p}$~\fi}
\newcommand{\RS}{{\rm R}\ifmmode_{\sun}\else$_{\sun}$~\fi}
\newcommand{\LS}{{\rm L}\ifmmode_{\sun}\else$_{\sun}$~\fi}
\newcommand{\MSbit}{{\rm M}\ifmmode_{\sun}\else$_{\sun}$\fi}
\newcommand{\RSbit}{{\rm R}\ifmmode_{\sun}\else$_{\sun}$\fi}
\newcommand{\LSbit}{{\rm L}\ifmmode_{\sun}\else$_{\sun}$\fi}
\maketitle

\begin{abstract}
We compiled data for 127 hosts (plus six candidates) and used them as constraints to construct interior models of the hosts using the {\small MESA} code. Two significant conclusions emerge from these models. First, except for a few stars, the hosts' metallicity ($Z_0$) is greater than 0.007. This suggests a possible suppression of the occurrence of planets below $Z_0\approx0.007$. Second, it concerns how chemical evolution unfolds in the galactic disc. For a given $Z_0$ value, considering the oldest stars, there is a linear relationship between $Z_0$ and age ($t_9$). This line is around $t_9=13.4$ Gyr at $Z_0=0$, a value consistent with the age of the Galaxy. The linear relationship continues until around $t_9=6$ Gyr, and the maximum value of $Z_0$ remains constant between $t_9=2-6$ Gyr. We further modelled 12 hosts classified as red clump (RC) stars in the literature, explicitly accounting for mass loss along the red giant branch. These models highlight the critical role of mass-loss assumptions in determining the initial masses and ages of RC hosts, and their implications for the survival and evolution of close-in planets. Another key outcome of this study is the discovery of the relationship between $Z_0$ and the observed metallicity ($Z_{\rm s}$) for the hosts. We obtain a useful expression for $Z_0$, the input parameter for the models, as a function of stellar mass, radius, and $Z_{\rm s}$. This expression can be used to estimate $Z_0$ based on the reduced surface metallicity due to microscopic diffusion. We also derive an expression for planetary mass relative to the orbital semimajor axis and host mass. This expression may indicate a mass distribution near the inner disc where these planets formed, except for hot-Jupiters. Planet radii appear to depend on the planet's mass and irradiation energy, as well as the orbital period.
\end{abstract}




\section{Introduction}

Understanding and modeling the internal structures of stars is crucial to astrophysics. Although traditional methods like spectroscopy and photometry provide valuable insights into stars, they become less effective as we probe deeper into a star's interior. This is where asteroseismology emerges as a revolutionary technique. Asteroseismology, the study of seismic waves within stars, has become an essential tool in astrophysics, providing invaluable knowledge of stellar dynamics. Many stars display oscillatory behaviour. These oscillations lead to periodic variations in brightness as they expand and contract. For Sun and solar-like stars, oscillations are driven by convective processes that occur in their outer layers. 

Fortunately, solar-like oscillations have been detected in many host stars, primarily from the light curves provided by recent space missions such as ESA’s Convection, Rotation, and planetary Transits \citep[CoRoT;][]{2006ESASP1306...33B}, NASA’s Kepler/K2 \citep{2010Sci...327..977B,2014PASP..126..398H}, and the Transiting Exoplanet Survey Satellite  \citep[TESS;][]{2015JATIS...1a4003R}. These missions have provided high-quality data for both asteroseismology and exoplanet science, greatly strengthening the connection between the two fields and allowing them to advance together {\citep{2015EPJWC.10102005V,2016IAUFM..29B.620H}}. Asteroseismology provides precise constraints on the internal structure and evolutionary state of stars, which are essential for the characterization of planetary systems {\citep{2013ApJ...767..127H,2015MNRAS.452.2127S}}. Fundamental characteristics of exoplanets, such as their mass, radius, age, and atmospheric composition, critically depend on the accuracy of the host star parameters. The more accurately we can determine the properties of the host star, the better we can assess the parameters of the orbiting planets.

In this study, we compiled asteroseismic and non-asteroseismic data of 133 solar-like oscillating stars and their planets from the literature. 6 of them are planet candidate systems ({KOI-5/KIC 8554498, KOI-75/KIC 7199397, KOI-268/KIC 3425851, KOI-364/KIC 7296438,  KOI-974/KIC 9414417, and KOI-2640/KIC 9088780}). We did not remove these stars, but we listed them separately. We examine how asteroseismology plays a key role in studying planetary systems and its importance in enhancing our understanding of these stars. Similar studies are available in the literature \citep{2013ApJ...767..127H,2019MNRAS.490.1509K}. These studies are based on a detailed modelling of the hosts and host candidates.

At the beginning of this study, approximately 40 hosts were compiled from the literature. With newly published studies, this number first increased to 80, and then to 127. A recent study by \cite{{2025ApJS..279...31L}} lists asteroseismic data for 142 hosts and computes fundamental properties of hosts and then planets using the asteroseismic scaling relations. Because the updates are endless, we had to stop at some point. 97 of the 142 hosts are available in our catalogue. In a similar study, \cite{2025AJ....170..212S} compiled an asteroseismic catalogue of 765 Kepler main-sequence (MS) and subgiant (SG) stars. Their analysis revealed 50 new detections, including seven planet candidate host stars. The catalogue includes 101 stars with planets. 38 of these stars are available in our catalogue.


The hosts in which planetary systems form and the extent to which planets' properties depend on their host mass, for example, are key questions. How planets evolve, both structurally and orbitally, is also a key question to answer.


In particular, determining planetary radii with high precision is of great importance. One of the main reasons for this is that the radius is the parameter that best represents the structure and evolution of planets compared to stars. After a planet forms, it cools relatively rapidly, reducing its radius. Planets with the biggest radii are those exposed to high irradiation (gas giants) {\citep{2000ApJ...534L..97B,2016ApJ...818....4L}}. Whether these planets cool over time can be understood from the change in their radius. Therefore, understanding how irradiation affects planetary evolution depends on accurately determining planetary radii. Moreover, even small uncertainties in the stellar radius can significantly affect the estimated planetary radius and, indirectly, our interpretation of the planet’s composition \citep{2018arXiv180402214L}. Thanks to asteroseismology, stellar radii can be determined with remarkable precision, thereby overcoming this issue.  

Equally importantly, the age of the system plays a key role, as it reveals how these planets have evolved over time. Since age is not directly observable, it remains difficult to constrain. Assuming that planets and their host stars form simultaneously, the planetary age can be inferred from the stellar age. The age of a planetary system provides important insight into the dynamical processes responsible for shaping the current structure of planetary systems, such as migration and star–planet interaction. It also plays a key role in the development of search strategies for potentially habitable planets \citep{2016IJAsB..15...93S}.
In this study, we compiled 127 planet-host stars  (along with six candidates) from the literature and constructed interior models with MESA \citep{Paxton2011, Paxton2013, Paxton2015, Paxton2018, Paxton2019, Jermyn2023}.


The study of planet-host stars that exhibit asteroseismic oscillations is fundamental for future stellar and exoplanet research. The primary aim of this study is to investigate planetary systems as a whole using asteroseismic data and to derive stellar and planetary parameters with higher reliability compared to other available methods. In this context, the study is intended to serve as a solid reference for the investigation of planetary systems and to support future research by addressing a wide range of questions related to both stars and their planets.

The paper is organized as follows: Section 2 presents the observed properties of the host stars and their planetary systems, together with the asteroseismic scaling relations. Section 3 describes the modelling methodology adopted in this study. The results of the computations are presented and discussed in Section 4. Finally, Section 5 provides the main conclusions of this study.

\section[]{Asteroseismic and Non-Asteroseismic Observed properties of planet hosts and asteroseismic scaling relations between them}
\begin{figure}	
\includegraphics[width=1.15\linewidth]{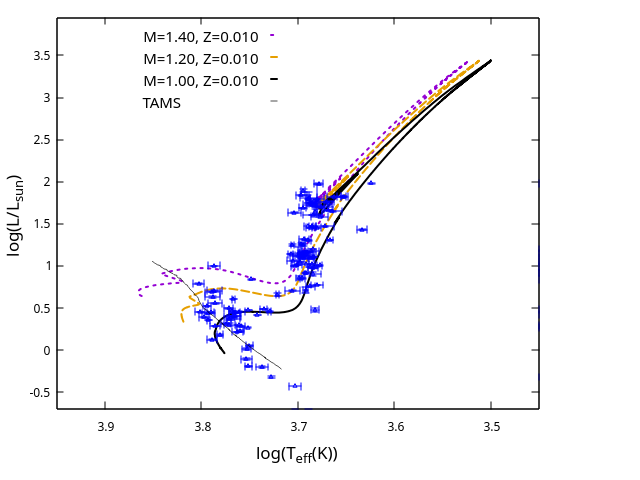}
    \caption{The classical HRD of the oscillating host stars. The thick solid and dashed lines represent the evolutionary tracks at $Z = 0.0100$ for 1.0, 1.2, and 1.4~$\rm M_{\sun}$. The thin solid line denotes the TAMS.}
    \label{fig:HRD_figure1}
\end{figure}
 The search for life on planets other than Earth, a highly popular field of research today, has led to a much more detailed study of planetary systems. In this study, knowledge of host stars, in particular, facilitates the study of planets. Accurate determination of stellar parameters such as radius, mass, age, and metallicity enables planetary radius and mass (see Table \ref{table:model}) with comparable accuracy. {This is because indirect detection methods such as the transit and radial velocity techniques derive planetary properties through their host stars. In the transit method, the planet-to-star radius ratio is measured from the transit depth, making the planetary radius directly proportional to the stellar radius. Similarly, radial velocity observations yield the planetary mass relative to the stellar mass, so the inferred planetary mass depends on the adopted stellar mass. Consequently, uncertainties in stellar parameters propagate directly into the derived planetary properties \citep{2018ASSP...49..119H}.}
 For this reason, asteroseismology, the most effective method for explaining the internal structure and evolution of stars, has become an essential tool in the study of planetary systems. We compiled asteroseismic and spectroscopic data of systems containing planets from the literature (Table \ref{table:alldata_obs}).
 
\subsection{Asteroseismic scaling relations}
Two fundamental asteroseismic quantities are significant: the value of the large frequency separation between oscillation frequencies (${\Delta \nu}$), and the frequency of maximum amplitude ($\nu_{\rm max}$). ${\Delta \nu}$ is defined as the frequency difference between modes of equal spherical degree $\ell$, and consecutive radial orders $n$. This quantity is intrinsically linked to the acoustic travel time within the star and is approximately proportional to the square root of the mean stellar density ($\bar{\rho}$) \citep{1980ApJS...43..469T,1993ASPC...42..347C}. $\nu_{\rm max}$ scales approximately as $g/\sqrt{T_{\rm eff}}$, where $g$ is the surface gravity \citep{Brown1991,Kjebed1995}. As stars evolve and surface gravity decreases, $\nu_{\rm max}$ shifts toward lower frequencies.

The mean of $\Dnu$ ($\braket{\Dnu}$) and $\nu_{\rm max}$  play an important role in determining the mass ($M$) and radius ($R$) of oscillating stars. 
The standard asteroseismic scaling relations, which allow the estimation of the radius ($R_{\rm sca}$) and mass ($M_{\rm sca}$) of an oscillating star, are given as follows: 
\begin{equation}
\label{eq:clasicsca}
\frac{R_{\rm sca}}{\rm R_{\odot}}=\frac{\numax/\nu_{\rm max\odot}}{(\braket{\Delta \nu}/\braket{\Delta \nu_\odot})^2}\left( \frac{T_{\rm eff}}{\rm T_{\rm eff\odot}}\right)^{1/2},
\end{equation}
\begin{equation}
\label{eq:clasicsca2}
\frac{M_{\rm sca}}{\rm M_{\odot}}=\frac{(\numax/\nu_{\rm max\odot})^3}{(\braket{\Delta \nu}/\braket{\Delta \nu_\odot})^4}\left( \frac{T_{\rm eff}}{\rm T_{\rm eff\odot}}
\right)^{3/2}. 
\end{equation}
Here, ${\braket{\Dnu_{\sun}}}$ and $\nu_{\rm max\odot}$ correspond to the solar values and the adopted values of ${\braket{\Dnu_{\sun}}}=135.1$ $\mu$Hz and $\nu_{\rm max\odot}=3090$ $\mu$Hz \citep{Sharma}. 


The small frequency separation ($\delta\nu$) is another crucial asteroseismic parameter \citep{Christensen1988}. It is defined as the frequency difference between modes of consecutive radial order $n$ and with a degree difference of two. The $\delta\nu_{02}$, calculated from modes with $l$ = 0 and $l$ = 2, is highly sensitive to the conditions in the stellar core during the MS, making it a valuable age indicator. As a star evolves along the MS, $\delta\nu_{02}$ decreases, being about 15 $\mu$Hz at the zero-age main sequence (ZAMS) and around 5 $\mu$Hz near the terminal-age main sequence (TAMS). For the Sun, $\delta\nu_{02}$ has a value of 9.8~$\mu$Hz \citep{2016MNRAS.462.1577Y}, which indicates that it is roughly halfway through its MS lifetime.

In this study, we analyse host stars at different evolutionary stages.
Only a few of the solar-like oscillating host stars in the literature are MS stars. The vast majority are SG and red giant branch (RGB) or red clump (RC) stars. Both the different structures of stars at different evolutionary stages and the fact that non-asteroseismic observational data, especially, have different uncertainty levels prevent us from applying a single modelling method. Therefore, it would be more useful to apply the technique based on each star's observational data. We used $\Dnu$ and surface metallicity ($Z_{\rm s}$) constraints for each star. We also used the reference frequencies ($\nu_{\rm min0}$ and $\nu_{\rm min1}$), effective temperature ($T_{\rm eff}$), luminosity ($L$), and $R$ calculated from the scaling relations as constraints according to the uncertainty level in the data of the stars. Analyzing the results obtained by applying several methods to some hosts would help us decide which method we prefer.

Detailed models of host stars are essential for many reasons: (i) good determination of the mass and radius of the stars allows good determination of the mass ($M_{\rm p}$) and radius ($R_{\rm p}$) of the transiting planets, (ii) the initial chemical composition of the star affects its own structure and also represents the chemical composition of the environment in which the planet formed, (iii) the age, which we can determine more precisely, especially with asteroseismic data, also reveals how the planets have evolved from the time they formed to the present.
\subsection[]{Non-asteroseismic properties of the planetary systems}
The luminosities of the stars in Table A1 are primarily obtained from references using GAIA DR3 and DR2 data \citep{2018A&A...616A...1G, 2021A&A...649A...1G}. Many studies provide {observed metallicity ([M/H])} and ${T_{\rm eff}}$ values for many of the hosts. Generally, these values are quite scattered. In this case, we tried to construct interior models for a host and obtain the observational data that best fit its model. The Hertzsprung–Russell diagram (HRD) plotted from the compiled data is shown in Fig. \ref{fig:HRD_figure1}. According to the TAMS line \citep{2015RAA....15.2244Y}, around 15 planet hosts are MS stars. Nearly 25 hosts are SGs, and the remaining hosts are RGB or RC stars. There may be a small number of core-helium-burning (CHeB) stars among the stars in the red giant (RG). Masses of the hosts range from 0.76 to 2.66 $\MS$. The radius interval is about 0.75–16.21 $\RS$.

[M/H] (or [Fe/H] ) and ${T_{\rm eff}}$ of the hosts are adopted from the spectroscopic studies in the literature (see Table \ref{table:alldata_obs}). [M/H] represents the metallicity of the photosphere of the hosts. For the interior models, however, the initial metallicity ($Z_0$) is the input parameter. For the evolved stars, $Z_0$ can be taken as $Z_{\rm s}$. However, for the hosts on and around MS, $Z_{\rm s}$ may significantly be less than $Z_0$, due to microscopic diffusion, depending on the age and depth of the stellar convective envelope. 
The $Z_{\rm s}$ of the planet hosts are computed from the [M/H] values:
\begin{equation}
\label{eq:ZsFeH}
Z_{\rm s}=10^{\rm [M/H]} Z_{\sun}, \nonumber
\end{equation}
where $Z_{\sun}$ is the solar metallicity and taken as 0.0134 \citep{Asplund2009}. 
For two of the hosts, namely KIC   6278762 and TIC 160224839, $Z_{\rm s}$ is small, $\sim$ 0.006. For most of the stars, {[M/H]} ranges from $-0.3$ to 0.3 dex. This implies that [M/H] for most of the systems varies between 0.007 and 0.027.  


As part of our study, we examined 133 planetary systems.  We compiled the masses, radii, orbital periods, and semimajor axis lengths of the planets from the literature to create a catalogue. The references for the planetary parameters in Table~\ref{table:alldata_obs} are marked in bold for clarity. In our sample, planetary masses range from 22.900 ${\rm M_J}$ to 0.0076 ${\rm M_J}$, while radii span 1.36 ${\rm R_J}$ to 0.027 ${\rm R_J}$. A total of 27 multiplanet systems are identified. Since asteroseismology allows us to determine the fundamental parameters of host stars with high precision, the properties of their planets can also be constrained with great accuracy. Therefore, these systems are of particular importance for studying planetary evolution and characterizing their fundamental properties (see Sections~\ref{sec:4.9} and~\ref{sec:4.10} ).

\begin{figure}
\includegraphics[width=1.15\linewidth]{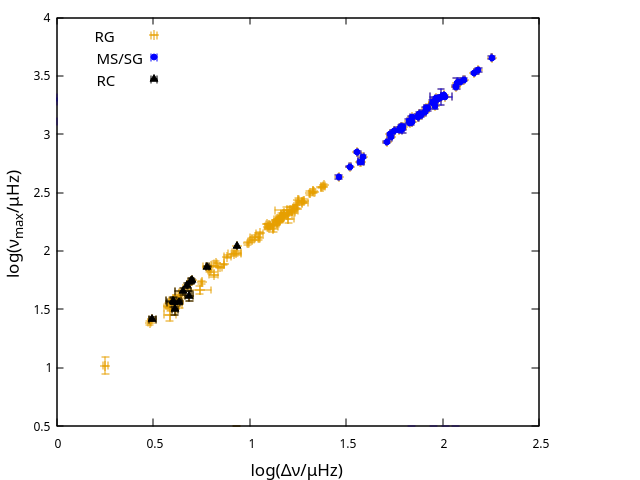}
\includegraphics[width=1.15\linewidth]{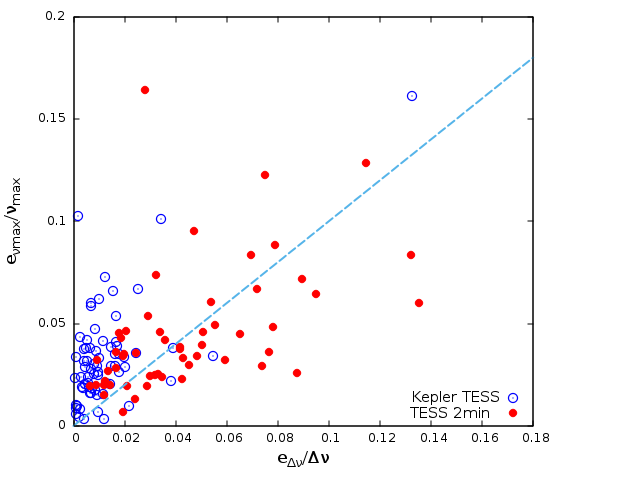}
    	
    \caption{ a) {$\numax$ is plotted  \wrt $\Dnu$ in logarithmic scales. The cross, filled circle, and triangle are for RG, MS/SG, and RC stars, respectively.} b) Uncertainty of $\numax$ is plotted \wrt $\Dnu$.}
    \label{fig:enumax}
\end{figure}

\subsection[]{Asteroseismic properties of the hosts}
The asteroseismic properties of the majority of planet hosts have been published in several studies \citep{2012A&A...543A..54A,2013ApJ...767..127H,2015MNRAS.452.2127S,2016MNRAS.456.2183D,2019MNRAS.490.1509K,2024ApJS..271...17Z}. {The quality of observational data of each planetary system and the number of observables are quite different from the others. While for some stars the available asteroseismic data include only their $\Dnu$, some of them 
have individual oscillation frequencies for many modes. Determining many oscillation frequencies is important because we can then use the reference frequencies as constraints.}
The asteroseismic properties of the Kepler RC and RGB targets are given in the APOKASC-2 catalogues. $\numax$ and $\Dnu$ of the TESS targets are mainly published in references. $\numax$ of the hosts is plotted with respect to $\Dnu$ in Fig. \ref{fig:enumax}(a). $\Dnu$ of RC and RGB stars ranges 2.5–8.8 and 0.45–18.7 $\mu$Hz, respectively.  $\numax$ of RC and RGB stars range 18.6–115 and 2.02–246.7 $\mu$Hz, respectively. The masses of the stars in the APOKASC-2 catalogue range between 0.8 and 2.5 ${\rm M_\odot}$ with a peak at around 1.2 ${\rm M_\odot}$. The stellar radii span from 2 to 40 ${\rm R_\odot}$.

 Fig. \ref{fig:enumax}(b) shows the relative uncertainty of $\numax$ ($e_{\numax}/\numax$) plotted against the relative uncertainty of $\Dnu$ ($e_{\Dnu}/\Dnu$). For most host stars, $e_{\numax}/\numax$ and $e_{\Dnu}/\Dnu$ are less than 3 per cent. More precise asteroseismic parameters lead to a more accurate determination of stellar mass. According to the standard error propagation of the scaling relations, the relative uncertainty in mass mainly depends on three times the fractional uncertainty in $\nu_{\max}$  and four times that in $\Delta\nu$. Therefore, reducing the uncertainties in these parameters improves the precision of the mass estimate and, consequently, allows for a more reliable determination of the stellar age.

During the RGB phase, stars expand significantly and lose mass with cold stellar winds; after reaching their maximum radii at the tip of the RGB, they undergo a helium flash and enter the RC phase. While the radii of RGB stars can reach values of $\sim 50\,R_\odot$, RC stars typically have radii in the range of $8$--$21\,R_\odot$. However, planet engulfment primarily occurs near the RGB tip, where the stellar radius and tidal interactions are strongest. A planet may either directly enter the stellar envelope or spiral into the star due to strong tidal interactions. This implies that planets located within $\sim 50$--$60\,R_\odot$ of their host stars are likely to be engulfed, unless they have already undergone a prior interaction.


\begin{figure}
\includegraphics[width=1.15\linewidth]{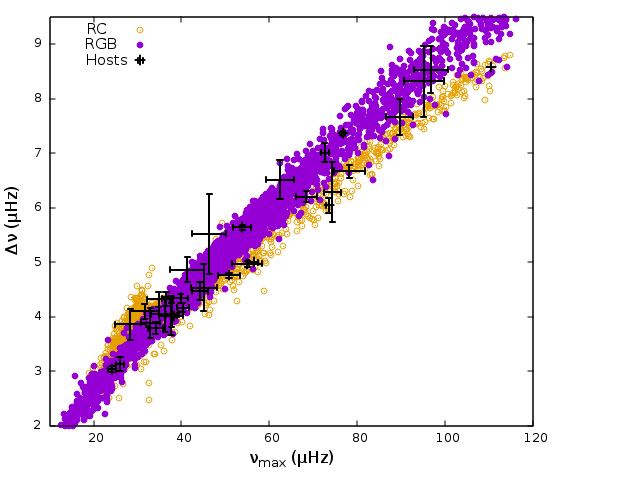} 	
    \caption{ $\Dnu$ is plotted \wrt $\numax$ for hosts to compare with RGB and RC stars in APOKASC-2. In this diagram, RC and RGB stars are slightly separated. Some of the hosts are likely RC stars. }
    \label{fig:Dnunumax}
\end{figure}

If a host is an RC star, then the period spacing ($\Delta \Pi_1$) is a significant constraint to model the interior of a star.
Unfortunately, $\Delta \Pi_1$ is not available for any host taking place in the RC region in Fig. \ref{fig:Dnunumax}. In Fig. \ref{fig:Dnunumax}, $\nu_{\rm max}$ is plotted against ${\Delta \nu}$ for the host stars, whose asteroseismic parameters are compiled from the literature, and compared with RGB and RC stars from the APOKASC-2 sample. The majority of the host stars are located near the RC sequence. 
Among these, 12 stars have been classified as RC objects in the literature. However, the classification of other stars located within the RC region remains uncertain and requires further investigation. 

The $\delta\nu_{02}$ values have been determined for 17 of the MS and SG stars, ranging from 4.3 to 9.0 $\mu$Hz. $\nu_{\rm min0}$ and $\nu_{\rm min1}$ obtained from the $\Delta\nu$–$\nu$ plot are determined for 23 of the 133 host stars. For 6 stars, $\nu_{\rm min0}$ or $\nu_{\rm min1}$ is detected. $\nu_{\rm min0}$ and $\nu_{\rm min1}$ have values in the ranges 245–4221 $\mu$Hz and 553–3849 $\mu$Hz, respectively (see Table \ref{table:alldata_obs}) .



\section[]{Method of modelling}
Instead of applying a single method to all stars during modelling, we used different methods based on the data in the literature and their uncertainties. These methods include MinZ, DZTR, DZTref, DZT, and DZTL. In addition, for 12 RC stars we applied two separate methods: DZLTeC and DZLTe3. {The method names indicate the observational constraints used in the modelling:}
\begin{enumerate}
    \item[(i)] {MinZ}: {`Min' denotes the unique solution corresponding to the minimum $\chi^2$, $Z_{\rm s}$.}

    \item[(ii)] {DZTR}: {$\Delta\nu$, $Z_{\rm s}$, $T_{\rm eff}$, $R$.}

    \item[(iii)] {DZT}: {$\Delta\nu$, $Z_{\rm s}$, $T_{\rm eff}$.}

    \item[(iv)] {DZTref}: {$\Delta\nu$, $Z_{\rm s}$, $T_{\rm eff}$, the reference frequencies.}

    \item[(v)] {DZTL}: {$\Delta\nu$, $Z_{\rm s}$, $T_{\rm eff}$, $L$.}

    \item[(vi)] {DZLTeC}: {$\Delta\nu$, $Z_{\rm s}$, $L$, $T_{\rm eff}$, the mass-loss coefficient ($\eta$) is determined from model calibration (eC).}

    \item[(vii)] {DZLTe3}: {$\Delta\nu$, $Z_{\rm s}$, $L$, $T_{\rm eff}$, the mass-loss coefficient is fixed at $\eta = 0.3$ (e3).}
\end{enumerate}

Among these methods, MinZ represents the most reliable approach and provides a unique solution. This method requires high-quality asteroseismic data, and stars with detected individual oscillation mode frequencies are the most advantageous stars. In this case, we know that the model we obtained by using $\Dnu$, reference frequencies, and $Z_{\rm s}$ as constraints is the unique model \citep{2025MNRAS.538..844O}.

In the DZTR method, the $\numax$ parameter is used when individually identified oscillation frequencies are available for stars. However, during the modelling of the stars, the seismic parameter that mainly affects the models and is used is $\Dnu$. The main reason for this is that the use of $\numax$ in these scaling relations has three important disadvantages, particularly for MS and SG stars.

First, we cannot correct the relation of $\numax$ to the fundamental parameters of the models. Secondly, the power spectra of hot F-type stars may exhibit more than one maximum, although this is not the case for all stars. In this study, there are two F-type stars with temperatures higher than 6000 K (KIC 9592705 and TIC 200723869). Third, the uncertainty in $\numax$ is generally higher than the uncertainty in $\Dnu$: $e_{\numax}/\numax = 6 e_{\Dnu}/\Dnu$. For the TESS 2-min cadence light curves \citep{2024ApJS..271...17Z}, $e_{\numax}/\numax$ and $e_{\Dnu}/\Dnu$ are high and close to each other. When we compare TESS and Kepler targets, we see that Kepler data have less uncertainty. For Kepler target stars, $e_{\Dnu}/\Dnu < 0.05$ and $e_{\numax}/\numax < 0.02$ . As a consequence, the stellar mass derived from the scaling relations becomes highly uncertain, because, as shown in equations (\ref{eq:clasicsca}) and (\ref{eq:clasicsca2}), the stellar radius depends on $\numax$ whereas the stellar mass depends on $\numax^{3}$. Therefore, the uncertainty in mass is significantly larger than that in radius.

This study is based on high-quality spectroscopic and asteroseismic data. For a large fraction of the sample, the uncertainty in $\Delta\nu$ is below $2$ per cent, 
while for most stars the uncertainty in $T_{\mathrm{eff}}$ is smaller than $100\,\mathrm{K}$. For these stars, $Y_0$ and $\alpha$ can be determined through the calibration of stellar models under observational constraints. This approach applies to most host stars with available reference frequencies and is referred to as the DZTref method.

If the oscillation frequencies of the individual modes are not available in the literature, we compute $\Dnu$, $Z_{\rm s}$, and $T_{\rm eff}$ as the observational constraints (Method DZT). 

For some stars, the spectroscopic $T_{\rm eff}$ is quite uncertain. In such cases, we either use luminosity derived using GAIA data (Method DZTL) or radius computed from the scaling relations (Method DZTR). 

For Methods DZT, DZTL, and DZTR, we assume that the chemical enrichment on the galactic disc proceeded as an increase in $Y$ ($\Delta Y$) is twice the increase in $Z$ ($\Delta Z$): $\Delta Y=2\Delta Z$. The heavy-element abundance can be determined from the stellar spectra. However, in order to determine the helium abundance of a star, either the solar initial helium abundance is adopted, or the helium abundance is calculated by assuming a relation between helium and metal enrichment. Although values of $\Delta Y / \Delta Z$ between 1 and 5 have been reported in the literature, a value of 2 is most commonly adopted \citep{2007MNRAS.382.1516C, 2021MNRAS.501..383T}. The values obtained with the ANK\.I code \citep{1965CaJPh..43.1497E, 1997A&A...326..187Y} and MESA are 1.95 and 1.58, respectively; therefore, we adopt $\Delta Y / \Delta Z = 2$ in this study.

It is well known that the star loses a significant amount of mass as it climbs the RGB branch, approximately 0.1–0.3 $\MS$ \citep{2011A&A...529A.137L, 2012MNRAS.419.2077M}. In cases where there is such a mass-loss, the planet is greatly affected for two reasons: First, since the central body loses mass, there is a significant change in the planet's orbit. Second, some of the mass lost by the star can be directly transferred to the planet, and in this case, the structure of the planet can change significantly. Such processes are complex, and depending on how mass accretion occurs, different outcomes can result. In this study, the mass-loss rate was investigated using Reimers' approach: mass-loss rate is proportional to $\eta LR/M$ {\citep{1975MSRSL...8..369R}.} 

For clarity, the methods used in this study can be summarized as follows:
\begin{enumerate}
  \item MinZ: When individual oscillation frequencies are available, the MinZ method provides a unique stellar model. This method was applied to a single star (KIC 8866102).
  \item DZTR: In this approach, stellar radii were derived from scaling relations. A total of 13 stars are modelled using this method.
  \item DZTref: This method was used for stars with high-quality asteroseismic and spectroscopic data, for which reference frequencies were available. 19 stars are analysed using this method.
  \item DZT: For stars without reference frequencies in the literature and with consistent spectroscopic $T_{\rm eff}$ values, the DZT method was employed. The models of 46 stars are obtained using this method.
  \item DZTL: Since the spectroscopic $T_{\rm eff}$ values of some stars are highly uncertain, the luminosities were computed using Gaia data. This method was applied to 41 stars.
  \item DZLTe3: 10 RC stars are analysed using this method. The mass-loss coefficient $\eta$ is taken as 0.3. The mass of the hosts is computed from the scaling relation given below: 
 \begin{equation}
\frac{M_{\rm sca}}{\rm M_{\odot}}=\frac{(L/{\rm L_{\odot}})^{3/2}}{(\braket{\Delta \nu}/\braket{\Delta \nu_\odot})^2}\left( \frac{\rm T_{\rm eff\odot}}{T_{\rm eff}}
\right)^{6}. 
\end{equation}

  \item DZLTeC: This method is similar to DZLTe3 and is also applied to 2 RC stars. The mass-loss coefficient $\eta$ is found from calibration of models.
\end{enumerate}

After summarizing the methods, the modelling procedure for the host stars can be described as follows: First, we compute the fundamental stellar parameters using the modified scaling relations. Then, we compare these results with the model results with the closest values of $Z$ and $M$ in the HRD.  
If there is a good agreement, then this model is the best model for the host star. If not, we test other M and $Z$ combinations.

The chemical composition of a star is among the most influential initial essential properties. Metallicity, in particular, significantly affects the structure of the nuclear core and the outer regions. Therefore, it is one of the most important ingredients of the interior models, after the stellar mass. In all the methods for asteroseismic modelling, we first use $Z_0$ calculated from the observed metallicity of the hosts. According to the difference between $Z_0$ and $Z_{\rm s}$, we estimate a new $Z_0$. The initial model value of $Z_0$ is taken to be approximately  $10$ per cent higher than $Z_{\rm s}$  , since some of these stars are in the middle and late stages of their MS evolution. An iterative procedure is then applied until consistency between $Z_0$ and $Z_{\rm s}$ is achieved.
For the evolved RGBs, $Z_0$ is close to $Z_{\rm s}$.

\subsection{Properties of MESA evolution code}

Nonrotating interior models of the planet hosts are constructed using {\small MESA} code version r23.05.1. Element diffusion is applied using the method proposed by \citep{1986ApJS...61..177P} for the stars with M<1.4 $\MS$. For the higher masses, the models are without diffusion. The standard mixing length theory proposed by \citet{Bohm1958} is used for the convection. The opacity is computed using {\small OPAL} tables \citep{Iglesias1993, Iglesias1996} at high temperature and the tables of \cite{Ferguson2005} at low temperature. Depending on the current state of evolution of the stars, the internal structure models cover pre-MS, MS, SG, and RGB phases.

In MESA, different equations of state (EOS) are blended depending on the temperature–density regime. For the stars studied here (MS, SG, RGB, and RC phases), most of the stellar models lie within the regime covered by FreeEOS\footnote{\url{https://freeeos.sourceforge.net/}}.

The adiabatic oscillation frequencies of the fitted model are computed using the ADIPLS package \citep{Christensen2008}. To account for the surface effect, the correction proposed by \cite{2008ApJ...683L.175K} is applied.
%
%
%
%
%
\section{Results and Discussions}
\subsection{Model properties across the observed properties}
\begin{figure}	
\includegraphics[width=1.15\linewidth]{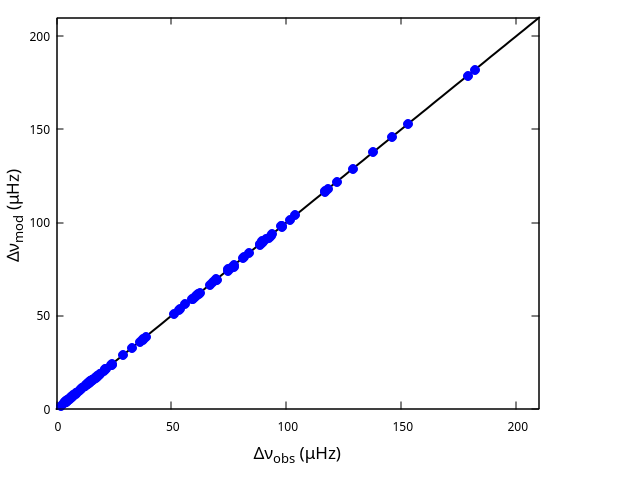}
\includegraphics[width=1.17\linewidth]{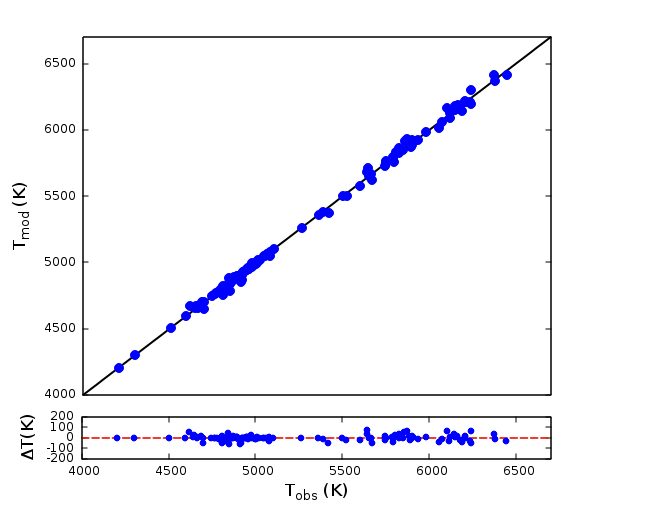}
    \caption{a) $\Dnu_{\rm mod}$ is plotted with respect to $\Dnu_{\rm obs}$. For most of the hosts, the difference between $\Dnu_{\rm mod}$ and $\Dnu_{\rm obs}$ is less than 1 per cent for most of the hosts. b) $T_{\rm mod}$ is plotted with respect to $T_{\rm obs}$.  The difference between $T_{\rm mod}$ and $T_{\rm obs}$ is less than 50 K for most of the hosts. }
    \label{fig: mod-obs}
\end{figure}

In Fig. \ref{fig: mod-obs}(a), the observed large frequency separation ($\Dnu_{\rm obs}$) of the host stars is compared with the model values ($\Dnu_{\rm mod}$). $\Dnu_{\rm mod}$ shows an excellent match with $\Dnu_{\rm obs}$, confirming the reliability of the modelling.

Fig. \ref{fig: mod-obs}(b) shows the comparison between observed ($T_{\rm obs}$) and modelled effective temperatures ($T_{\rm mod}$). Overall, $T_{\rm obs}$ and $T_{\rm mod}$ are in perfect agreement, although some scatter is visible. This scatter is likely due to the relatively significant uncertainties in the observed stellar temperatures.

\subsection{Planet hosts in the RC evolutionary phase}
\begin{table}
    \centering

\caption{The interior models of 12 RC stars are constructed with mass loss. The mass is mostly lost around the RGB tip. Initial and final masses are listed in the second and third columns, respectively. The Reimer coefficient and the amount of lost mass are given in the third and fourth columns, respectively. } 
\begin{tabular}{llcccccccc}
\hline
   Host       & $M_{0}$ &      $M$  &   $\eta$ & $\Delta M$ \\ 
              & $\MS$   &     $\MS$ &          & $\MS$       \\
\hline
TIC  17554529 &    2.395 &    2.387 &    0.300 &   -0.008 \\
TIC  28763463 &    1.754 &    1.684 &    0.300 &   -0.071 \\
TIC  49430557 &    1.794 &    1.725 &    0.300 &   -0.069 \\
TIC 129649472 &    1.174 &    1.036 &    0.300 &   -0.138  \\
TIC 160224839 &    0.974 &    0.814 &    0.300 &   -0.160 \\
TIC 232074535 &    1.952 &    1.901 &    0.300 &   -0.050 \\
TIC 246937460 &    1.196 &    1.074 &    0.300 &   -0.122 \\
TIC 247119634 &    1.822 &    1.750 &    0.300 &   -0.072 \\
TIC 257005016 &    1.666 &    1.556 &    0.420 &   -0.109 \\
TIC 284181945 &    2.227 &    2.185 &    0.580 &   -0.042 \\
TIC 354489950 &    2.390 &    2.385 &    0.300 &   -0.005 \\
TIC 441797545 &    1.410 &    1.314 &    0.300 &   -0.096 \\
\hline
\end{tabular}
\label{table:mod_RC_DM}
\end{table}
A star that has passed the RGB tip and reached the RC region has experienced significant mass-loss. Even if we can accurately determine the present-day M and R of an RC star from modelling or scaling relationships, determining the amount of mass-loss is crucial because age largely depends on the initial mass. The coefficient $\eta$ is taken as 0.3 for the 10 hosts. For the two hosts, namely TIC 257005016 and TIC 284181945, we obtain the best models with $\eta$=0.42 and 0.58, respectively. The basic properties of the models are listed in Table \ref{table:mod_RC_DM}.

\begin{figure}
	\includegraphics[width=1.1\linewidth]{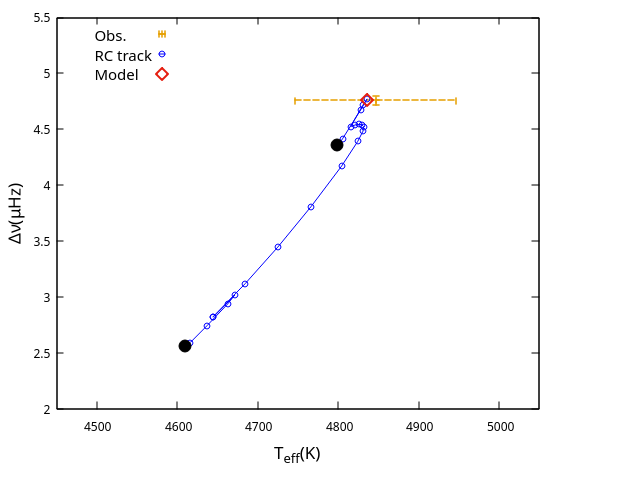}
    \includegraphics[width=1.1\linewidth]{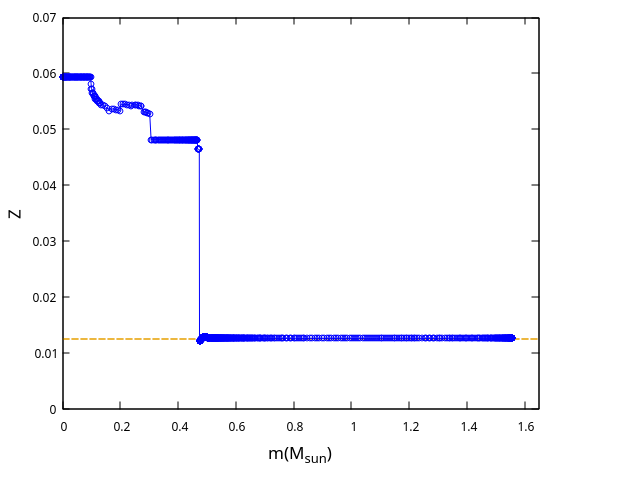}
    \includegraphics[width=1.1\linewidth]{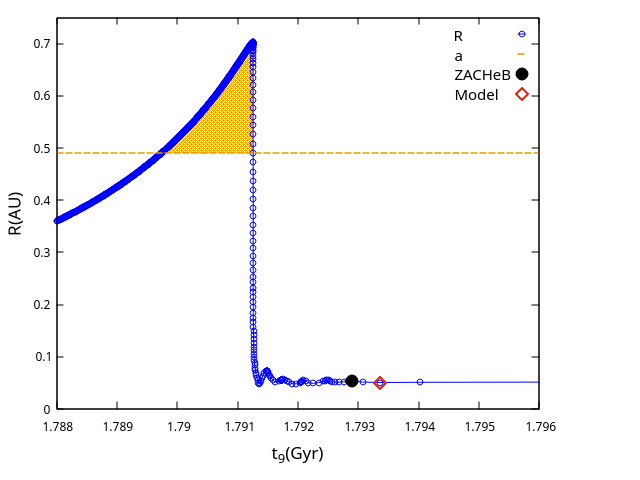}
    \caption{a) Evolutionary track of TIC 257005016 is plotted in the $\Dnu$-$\teff$ graph. b) Z throughout the interior model is plotted with respect to m. In the core region, Z is significantly greater than $Z_0$, where the triple alpha process produces $^{12}$C. c) TIC 257005016’s age versus radius. The dashed lines indicate the distance of the planet from its host star. The planet TIC 257005016b may be a diving planet. }
    \label{fig:RChostTIC257005016}
\end{figure}

In Fig. \ref{fig:RChostTIC257005016}, details of the RC model of TIC 257005016 are shown. In the upper panel, $\Dnu$ is plotted with respect to $\teff$ from the zero-age core helium burning (ZACHeB) to the terminal-age core helium burning (TACHeB). The host initially rises slightly and then descends considerably. Observations only align with this when it is at the peak of its evolutionary trace. The middle panel shows the host's mass versus Z. Although the Z is very close to the initial value of $Z_0$ in the outer parts, it rises to 0.06 in the core as $^4$He is converted to $^{12}$C. Interestingly, the helium flash leaves a relatively irregular Z distribution in the core. 

\subsection{Are diving (dalg\i \c{c}) planets real?}
The bottom panel of Fig. \ref{fig:RChostTIC257005016} shows the time variation of TIC 257005016's radius near the RGB tip ($R_{\rm tip}$). This star's radius grows to 0.7 AU, whereas TIC 257005016b's $a$ is 0.49 AU. The host's $R$ is greater than $a$ for approximately 1 Myr, starting from 1.79 Gyr. How can this happen? Two more stars have $R_{\rm tip}$ values exceeding the $a$ of their planets. These stars are TIC 129649472 and TIC 160224839. How could the planet have survived? {The survival of planets around evolved stars has been widely discussed in the literature \citep{2007ApJ...661.1192V,2011ApJ...737...66K,2023ApJ...950..128O}. In particular, \cite{2023Natur.618..917H} discussed the survival of 8 UMi b (TIC 257005016), and proposed explanations for its survival. Considering both previous studies and our results,} we discuss the following possible scenarios: 
(i) The density of the outer layers around the RGB tip is very low; the planet may have dived and resurfaced. (ii) These planets may be second-generation planets formed in the disc created by the mass lost by the host. (iii) If the planet gained some of the mass lost by the host, it may have been dragged towards the host as $M_{\rm p}$ increases and its $a$ decreases. (iv) The host's initial mass may be much greater, in which case $R_{\rm tip} > a$ would not be true since $R_{\rm tip}$ is lower. Investigating planetary hosts in the RC phase is of particular importance as it is also crucial for understanding the ultimate end of Earth.

\subsection{Thick disc members}
Three of the 127 hosts, namely KIC 6278762, TIC 136916387, and TIC 160224839, have [$\alpha$/Fe] values of 0.26, 0.15, and 0.19, respectively. The basic properties of the models are listed in Table \ref{table:mod_RC_DM} \citep{2015ApJ...799..170C,2025MNRAS.541.2459W,2016ApJ...817...40F}. Since the [$\alpha$/Fe] of these three stars is greater than 0.1 dex \citep{2021A&A...645A..85M}, these stars are thick disc members and among the oldest hosts (see Fig. \ref{fig:Zt0}). {For these stars, the observed [Fe/H] and [$\alpha$/Fe] values were used to calculate the metallicity [M/H] using the relation of \cite{1993ApJ...414..580S}. The corresponding $Z_{\rm s}$ values were then calculated from
[M/H]. Since these stars are evolved, we adopted $Z_0 \approx Z_{\rm s}$
and used this value as the metallicity input in the models.
These models were then used to determine $R_{\rm tip}$.  }

\subsection{Asteroseismic HR diagram}
The age of a star is a key parameter for understanding its internal structure and the evolution of stellar and planetary systems. Asteroseismology enables the determination of stellar ages with significantly higher precision than other methods. An asteroseismic HRD is an effective tool for this purpose \citep{1993ASPC...42..347C}.

\begin{figure}
	\includegraphics[width=1.\linewidth]{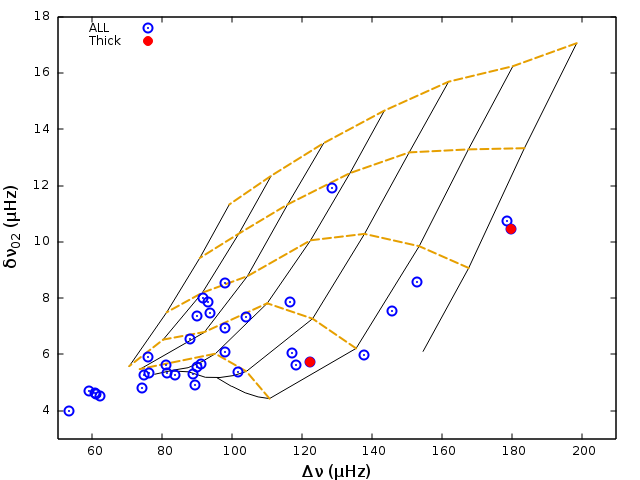}
    \caption{Asteroseismic HRD for models of the hosts. The grids are prepared for $M=0.8-1.4$ $\MS$ with steps of 0.1 $\MS$ and $Z_0=0.02$ (nearly vertical solid lines). The dashed lines show the evolutionary phases from ZAMS (at the top) to TAMS (at the bottom). The solid circles represent the thick stars KIC 6278762 and TIC 136916387.}
    \label{fig:seisHR}
\end{figure}

In Fig. \ref{fig:seisHR}, the $\Dnu$ is plotted against the small frequency separation to construct an asteroseismic HRD for the host star models. In this diagram, solid lines indicate stellar mass (1.4 $\MS$ on the left and 0.8 $\MS$ on the right), while dashed lines indicate the MS phases (ZAMS at the top and TAMS at the bottom). The least evolved star in the MS phase is KIC 4141376, with a $\delta \nu_{02}$ of 11.9 $\mu$Hz. Two more systems have not yet reached the halfway point of their MS lives; these are KIC 6278762 and KIC 8478994. These stars have $\delta \nu_{02}$ around 10.5 $\mu$Hz. Since the masses of both stars are less than 0.8 $\MS$, they are located to the right of the 0.8 $\MS$ line. The stars form two clear groups: one is concentrated near the TAMS, while the others appear to have completed about three-quarters of their MS lifetime.

\subsection{Metallicity--age relationship}
\label{sec: 4.6}
\begin{figure}	
\includegraphics[width=1.15\linewidth]{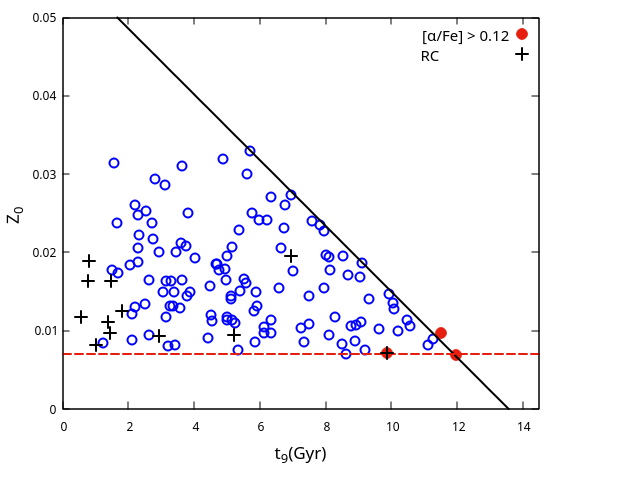}
\includegraphics[width=1.15\linewidth]{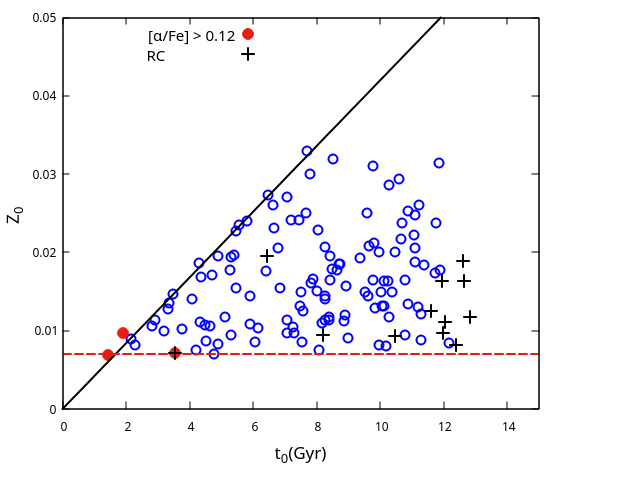}
\caption{ a) $Z_0$ as a function of stellar age ($t_9$). The solid line is the border for the stars with the highest metallicities of their ages, expressed as $Z=-0.00395t_9+0.053$. The dashed line indicates $Z_0 = 0.007$, above which all host stars in our sample are located. While the filled circle show the thick disc members ($[\alpha/\mathrm{M}] > 0.12$), the cross is for the RC stars. b) $Z_0$ as a function of $t_0$, defined as $t_0 = t_{\mathrm{MW}} - t_9$, where $t_{\mathrm{MW}} = 13.4$ Gyr.}
    \label{fig:Zt0}
\end{figure}

The $Z_0$--$t_9$ relationship obtained from host internal-structure models is shown in Fig. \ref{fig:Zt0}. This figure shows two interesting features. One of these is about galactic chemical evolution: For stars formed from gas clouds with the fastest increasing metallicity, there is a linear relationship between $Z_0$ and $t_0=t_{\rm MW}-t_9$. $t_{\rm MW}$ is taken as 13.4 Gyr \citep{2004A&A...426..651P}. 

Old stars formed in the early epochs of the Milky Way generally exhibit lower metallicities, while younger stars exhibit higher metallicities. This trend is consistent with expectations from galactic chemical evolution. Another critical point is that no planetary systems are found below a certain metallicity threshold. In our samples, we observe that host stars with $Z_0$ < 0.007 do not harbour planetary systems. This may be the main link between stellar metallicity and the occurrence of planetary systems. Hence, this aspect plays a crucial role in studying stellar-planetary evolution and in improving our understanding of planet formation processes.


Some host stars exhibit very low measured [M/H] ([Fe/H]) ratios, corresponding to low $Z_{\rm s}$ values \citep{{2007ApJ...665.1407C},{2023AJ....166...49D}}. This arises from the difference between $Z_{\rm 0}$ and $Z_{\rm s}$. In particular, stars evolving along the MS exhibit low $Z_{\rm s}$. After the MS, however, as the convective envelope deepens and previously buried heavy elements are transported back to the surface, $Z_{\rm s}$ approaches $Z_{\rm 0}$ (see Appendix \ref{Appendix B}  for further details).  

\subsection{The initial metallicity versus surface metallicity}
The initial metallicity plays a key role in determining the structure of stars. Because more energy is expended on ionization when the number of heavy elements is high, less of the energy gained from the collapse during star formation is spent on heating the deep interior. In this case, the temperature increases less, leading to slower nuclear reactions and lower luminosity. From this perspective, the initial $Z$ has a significant impact on the structure of stars (especially on their luminosity). $Z$, determined from spectral observations, is actually the surface heavy element abundance $Z_{\rm s}$. Structure and evolution, in turn, depend primarily on $Z_0$. The crucial question here is whether we can calculate $Z_0$ from $Z_{\rm s}$.

\begin{figure}	
\includegraphics[width=1.15\linewidth]{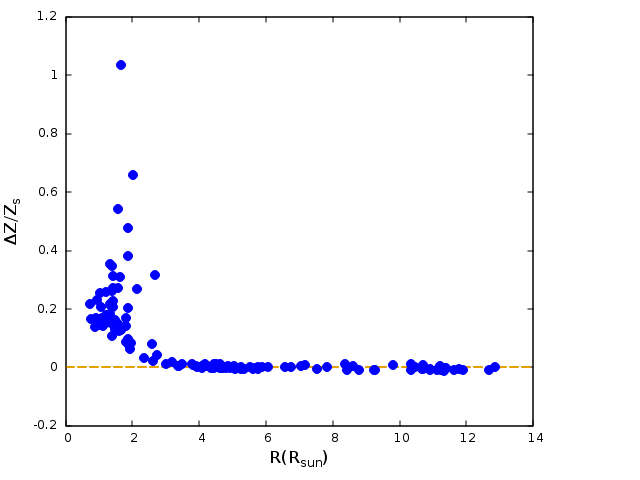}
\includegraphics[width=1.15\linewidth]{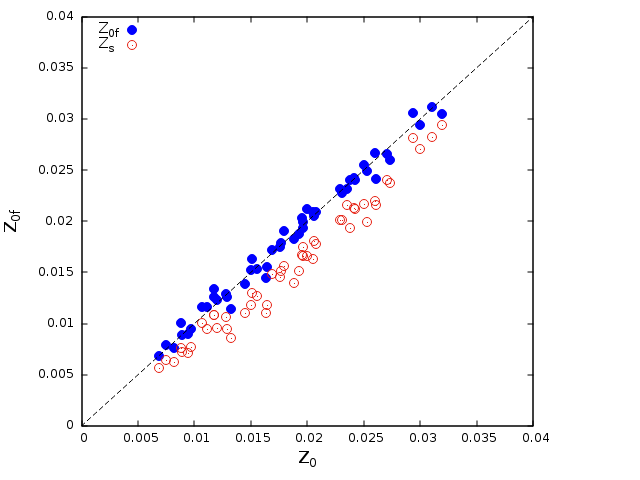}
    \caption{a) The fractional difference between $Z_0$ and $Z_{\rm s}$ is plotted with respect to radius for the interior models of the hosts. The greatest difference is for KIC 11401755. While its $Z_0$ is 0.011, its $Z_{\rm s}$ is about half of $Z_0$. The second greatest difference occurs for KIC 10666592. The difference is due to microscopic diffusion and is significant for the stars with $R<3 {\rm R_{\odot}}$. Assuming the difference is a function of $M$, $Z_{\rm s}$, and $R$, we obtain a fitting formula for these stars, except KIC 10666592 and KIC 11401755. b) $Z_{\rm 0f}$ (filled circles), computed from the fitting formula, is plotted against the original $Z_0$. Also plotted is the $Z_{\rm s}$ (circles).}
    \label{fig: Z0fZ0_figure}
\end{figure}
Using the data of host interior models, the fractional difference $\Delta Z/Z_{\rm s}$ ($(Z_{0}-Z_{s})/Z_{s}$) is plotted against $R$ in Fig. \ref{fig: Z0fZ0_figure}a. The difference is particularly significant for the range $R<3$ \RSbit. For larger radii, the deepening convective envelope eliminates the chemical separation caused by microscopic diffusion. For $R < 3$ \RSbit, $R$ roughly represents the evolutionary state. Diffusion effect depends on the depth of the convective envelope. The depth of the convective envelope depends on $M$ and $Z$. Then we can consider $\delta Z/Z_{\rm s}$ as a function of $M$, $Z_{\rm s}$, and $R$: $f(M,Z_{s},R)$. We assume a function as:
\begin{equation}
\label{eq:eq4}
  \frac{\delta Z}{Z_{s}}=f(M,Z_{s},R)=b \frac{M}{M_{\sun}}+cZ_{s}+d\frac{R}{R_{\sun}}.
\end{equation}
From this function, we obtain the fitting formula using the data of 50 hosts for which $dZ>0.0005$.
Then, we can compute $Z_0$ in terms of $M$, $Z_{s}$, and $R$:
\begin{equation}
Z_{0\rm f}=Z_{\rm s}(1+f)
\end{equation}
In Fig. \ref{fig: Z0fZ0_figure}(b), $Z_{0{\rm f}}$ is plotted with respect to $Z_0$. The metallicities calculated from the model and the $Z_0$ relationship are in excellent agreement.

In our host sample, there are a few stars in the vicinity of the ZAMS. Therefore, equation (\ref{eq:eq4}) is only valid for evolved MS and SG stars. For the young stars, $\delta Z/Z_{\rm s}$ is a double-valued function of $R$.

\subsection{Comparison of stellar radius and mass with the literature}
\begin{figure}	
\includegraphics[width=1.15\linewidth]{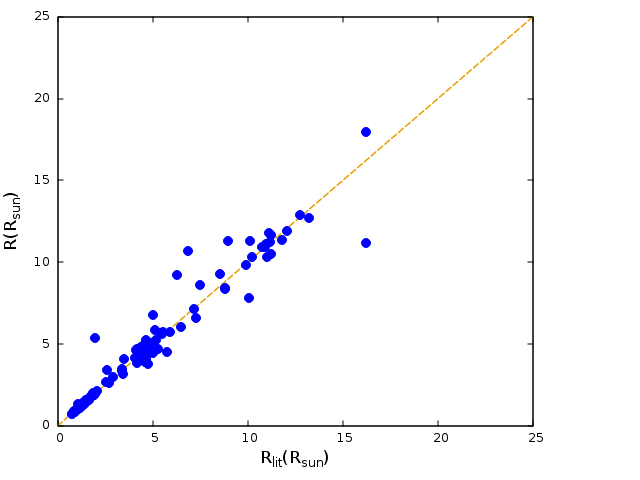}
\includegraphics[width=1.15\linewidth]{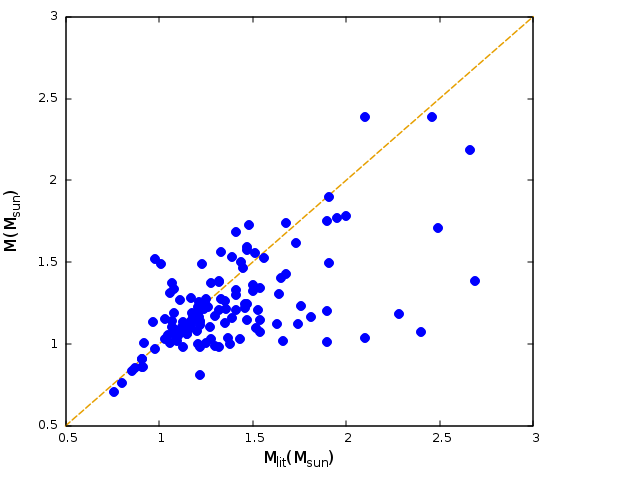}
    \caption{ a)  $R$ is plotted with respect to $R_{\rm lit}$ for the 127 planet hosts. The radii are in good agreement with the literature. b) $M$ is plotted with respect to $M_{\rm lit}$. In contrast, the masses show a larger scatter.}
    \label{fig:RM_kiyas_figure}
\end{figure}
In Figs. \ref{fig:RM_kiyas_figure}(a) and (b), the radii and masses of the models are compared with the radii ($R_{\rm lit}$) and masses ($M_{\rm lit}$) from the literature, respectively. The scatter in the radii is significantly smaller than that in the masses. This is expected, since the radii can be more directly constrained from observational quantities such as luminosity (e.g. from Gaia) and effective temperature. In contrast, masses are usually derived from $\log g$ and $R$. Therefore, uncertainties in different observational parameters affect the mass estimates. In particular, the poor determination of $\log g$ leads to a larger scatter in the masses. 

The use of scaling relations without corrections, especially for RGB and SG stars, also contributes to the scatter in the derived masses and radii. {Previous studies have shown that the classical asteroseismic scaling relations for red giants do not exhibit a direct dependence on metallicity. However, the corrected asteroseismic scaling relations show a dependence on metallicity through both the   $\Delta\nu \propto \sqrt{\bar{\rho}}$  relation  \citep[Fig.~5:][]{2023MNRAS.518.5552Y} and the $\nu_{\rm max}\propto g/\sqrt{T_{\rm eff}}$ relation \citep[Eq.~11:][]{2023MNRAS.518.5552Y}. Applying these corrections significantly improves the agreement between asteroseismic and dynamical parameters. In particular, \cite{2026MNRAS.549g1069Y} showed that the corrected radius scaling relation ($R_{\rm sca}$) reproduces the dynamically measured radii of red giants in eclipsing binaries with excellent accuracy. The corrected mass scaling relation ($M_{\rm sca}$) likewise improves the agreement with dynamical masses, although a somewhat larger scatter remains. These studies cover a wide range of metallicities, including those of the stars in our sample.} 


In this comparison, we ensured that the $M_{\rm lit}$ and $R_{\rm lit}$ were taken from the same source that provides the planetary parameters. {It should be noted that the masses and radii adopted from the literature were not derived using a single homogeneous method. Most of the literature values were obtained from asteroseismic analyses \citep[e.g.][]{2013ApJ...767..127H}, whereas the remaining values were primarily derived through isochrone fitting or grid-based modelling using spectroscopic and photometric constraints \citep[e.g.][]{2016ApJS..225...32B,2017AJ....153...51W,2023PASJ...75.1030T}. Since these methods rely on different assumptions and have  different systematic uncertainties, part of the scatter observed in the mass and radius comparisons shown in Fig. \ref{fig:RM_kiyas_figure} is expected.}


\subsection{Revised radius and mass of the planets}
\label{sec:4.9}
\begin{figure}	
\includegraphics[width=1.1\linewidth]{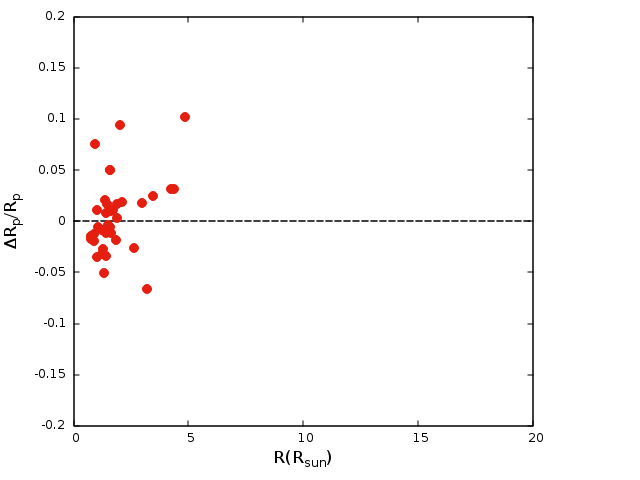}
\includegraphics[width=1.1\linewidth]{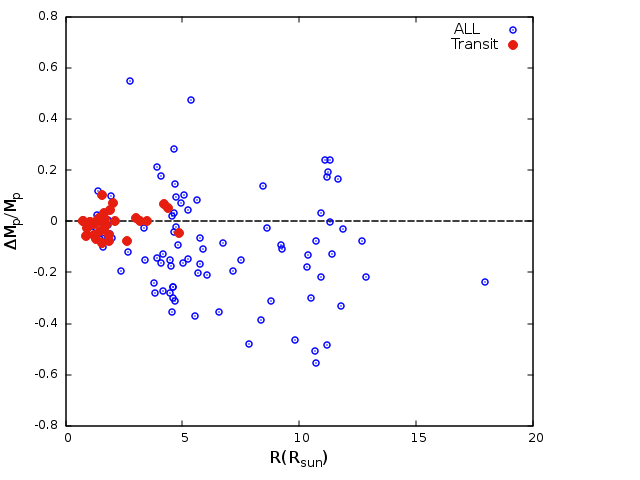}
\includegraphics[width=1.1\linewidth]{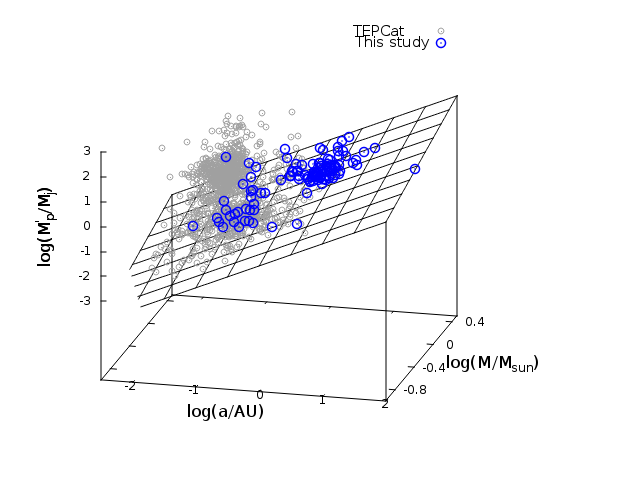}
    \caption{a)  $(R_{\rm p}'-R_{\rm p})/R_{\rm p}$  and b) $(M_{\rm p}'-M_{\rm p})/M_{\rm p}$ are plotted \wrt $R$ of the hosts. Radii of the transit planets are determined from the observations in units of their host radii (filled circles). The difference for $R_{\rm p}$ is less than 10 per cent for 37 transit planets. For these planets, the difference for $M_p$ is similar to that of $R_p$. However, for all of the planets (circle), the difference is less than 50 per cent. c) In the lower panel, $M_{\rm p}'$ is plotted with respect to $a$ and $M$ in logarithmic scales. }
    \label{fig:dRMp_figure}
\end{figure}

Due to two spheres passing in front of each other, the light curve of a planetary system changes. The transit depth, $\Delta F$, with $F$ defined as the total observed flux,
is given as \citep{2003ApJ...585.1038S} 
\begin{equation}
    \Delta F=\left(\frac{R_{\rm p}}{R}\right)^2.
\end{equation}
Since the radius ratio will be constant for a given value of $\Delta F$, we can also calculate the revised $R_{\rm p}$ ($R_{\rm p}'$) from the newly calculated $R$ of the host. $R_{\rm p}'$ is also listed in Table \ref{table:model}. The fractional difference between $R_{\rm p}'$ and $R_{\rm p}$ (($R_{\rm p}'-R_{\rm p})/R_{\rm p}$) is plotted with respect to $R$ in the upper panel of Fig. \ref{fig:dRMp_figure}. The difference is less than 10 per cent. 

The revised planetary mass is computed from the ratio of the new stellar mass to the old one. 
In the lower panel of Fig. \ref{fig:dRMp_figure}, the fractional difference between $M_{\rm p}'$ and $M_{\rm p}$ (($M_{\rm p}'-M_{\rm p})/M_{\rm p}$) is plotted against $R$. For most planets, the difference is quite large, reaching up to 50 per cent. This difference is due to the difference between the new stellar mass and the old mass. For TIC 200093173b, the difference is about 55\%. The mass of the host is given as $0.98\pm 0.125$ $\MS$ in \cite{2022ApJS..262...21F} and is found to be 1.517 $\MS$ in the present study.
It is worth noting that $M_{\rm p}$ here is mainly $M_{\rm p} \sin i$.
See Fig. \ref{fig:dRMp_dRM} in Appendix \ref{app:C} for the differences between revised and
literature values of radius and mass of the planets and the hosts.

\subsection{Expression for $M_{\rm p}'$ as a function of $a$ and $M$ }
\label{sec:4.10}
In Fig. \ref{fig:dRMp_figure}, $M_{\rm p}'$ is plotted with respect to $M$ and $a$ in logarithmic scales. The grids shows the fitted function $f(x,y)$:
\begin{equation}
\label{eq: eq7}
    \log\left(\frac{M_{\rm p}'}{\MJ}\right)= (1.10\pm0.35)\log\left(\frac{M}{\MS}\right) + (1.07\pm0.07)\log\left(\frac{a}{AU}\right).
\end{equation}
According to this relationship, the planetary mass increases with increasing $M$ and $a$. This result is interesting because it may indicate the density that existed under the conditions of the disc (presumably in the inner parts of the disc) that produced the planet with mass $M_{\rm p}'$. Other systems in the TEPCat catalogue \citep{2011MNRAS.417.2166S} are also plotted for comparison. We observe that there are many hot Jupiters in TEPCat with $M_{\rm p}' > $ 0.3 $\MJ$ and $a < $ 0.1 au. However, such hot Jupiters are relatively rare among the b planets of our 127 hosts. Hot Jupiters form a distinct group, and therefore equation (\ref{eq: eq7}) is not valid for them. 

{The fact that our sample does not contain many hot Jupiters around stars exhibiting solar-like oscillations is noteworthy.}
{Although it is widely accepted that hot Jupiters reach their close-in orbits through migration processes \citep{1996Natur.380..606L,2018ARA&A..56..175D}, tidal interactions during stellar evolution may lead to the engulfment of close-in giant planets by their host stars \citep{2009ApJ...705L..81V}. It has been suggested that this effect is more pronounced for higher-mass evolved stars ($M_{\star} > 1.5\,M_{\odot}$; \citealt{2011ApJ...737...66K}). However, the mass of most stars in our sample is below this threshold. Therefore, planetary engulfment alone is unlikely to explain the paucity  of hot Jupiters in our sample. Furthermore, Fig. \ref{fig:dRMp_figure} represents only a subset of our sample, as planetary masses are not available for many systems. If the masses of these planets are determined in future studies, some could be classified as hot Jupiters, potentially increasing the observed hot Jupiter population. Hence, whether selection effects contribute to the paucity  of observed hot Jupiters should be evaluated in more detail.}

In contrast, while there are many planets with $M_{\rm p}' > $ 1 $\MJ$ and $a >$ 1 au in this study, planets in this region are rarely seen in the TEPCat catalogue. The radii of the hosts of these planets exceed 3.5 $\RS$.


\subsection{How $R_{\rm p}$ depends on $M_{\rm p}$ and orbital parameters?}
It is well known that there are three distinct M--R relationships for planetary mass ranges: rocky, Neptune-like, and Jovian. For the latter two in particular, the M--R relationship exhibits significant scattering. The host's irradiation flux plays a significant role in this scattering. A much more precise relationship emerges when the irradiated energy per gram per second ($l_-$) is used instead of the flux \citep{{2014MNRAS.445.4395Y}}. In addition to these parameters,
$R'_{\rm p}$ shows a secondary dependence on the orbital period $P$. The fitting formula for $R'_{\rm p}$ ($R'_{\rm pfit}$) is obtained as
\begin{eqnarray}
\label{eq: eq8}
&&\log\left(\frac{R_{\rm pfit}'}{\RJ}\right)= (-1.39\pm0.20)+(0.300\pm0.039)\log\left(\frac{l_{-}}{l_0}\right)+ \nonumber \\
&&(0.411\pm0.027)\log\left(\frac{M_{\rm p}'}{\MJ}\right)+(0.420\pm0.067)\log\left(\frac{P}{\rm day}\right),
\end{eqnarray}
where $l_0$ is the energy received per unit mass and time by a planet with 1 $\MJ$ and 1 $\RJ$
at 1 au in our Solar System ($l_0=1.106 \times 10^{-4}$ erg g$^{-1}$ s$^{-1}$).

\begin{figure}	
\includegraphics[width=1.1\linewidth]{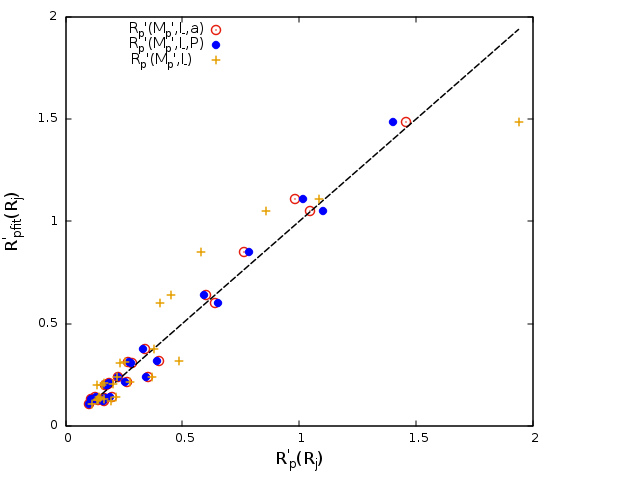}
\includegraphics[width=1.1\linewidth]{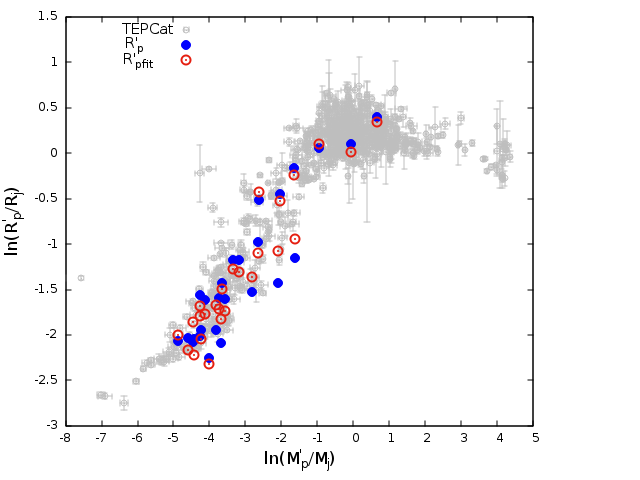}
    \caption{a) $R'_{\rm pfit}$ (filled circle) given in equation \ref{eq: eq8} is plotted with respect to $R'_{\rm p}$. We obtain another fit for $R'_{\rm p}$ by substituting $a$ (circle) for $P$ in equation \ref{eq: eq8}. The two fits give similar results. To see the direct effect of the orbital parameters, we obtain an expression as a function of $M'_{\rm p}$ and $l_{-}$ (+). The $R'_{\rm pfit}$ obtained from this expression is not in good agreement with $R'_{\rm p}$. b) $\ln(R'_{\rm p})$ (filled circle) and $\ln(R'_{\rm pfit})$ (circle) are plotted with respect to $\ln(M'_{\rm p})$. For comparison, the planets in TEPCat are also shown (gray errorbars).}
    \label{fig:RpMp_figure}
\end{figure}

In Fig. \ref{fig:RpMp_figure}, $R'_{\rm pfit}$ is plotted with respect to $R_{\rm p}'$. 
The average difference between $R'_{\rm pfit}$ and $R'_{\rm p}$ is approximately 14 per cent. The biggest difference is for KIC 8494142b, about 43 per cent. (The host star TIC 612908 was excluded from the plots in Fig. \ref{fig:RpMp_figure}, because this planet has a very large radius despite having low incident flux.)

Another notable feature in Fig. \ref{fig:RpMp_figure}(a) is that the agreement improves when $R_{\rm p}'$ is expressed as a function of the orbital parameters ($a$ and $P$). In this case, the data points show less scatter, with better agreement between $R'_{\rm pfit}$ and $R'_{\rm p}$.

\subsection{Notes on individual hosts}
\subsubsection{16 Cyg B (KIC 12069449)}
The asteroseismic data of 16 Cyg B is obtained from Kepler light curves \citep{2012ApJ...748L..10M}. Its individual oscillation frequencies are derived from this data. Its $\delta \nu_{02}$ is 6.6 $\mu$Hz and is an essential constraint on age indicators. While its $\nu_{\rm min0}$ is well defined, $\nu_{\rm min1}$ is around the first interval of the data (2044 $\mu$Hz). Detailed interior models of 16 Cyg A and B are constructed by \cite{2012ApJ...748L..10M} using different codes. M, R, and age are found to be 1.07$\pm$0.02 $\MS$, 1.127$\pm$0.007 $\RS$, and 6.7$\pm$0.4 Gyr, respectively. 
These models are in good agreement with the asteroseismic and non-asteroseismic data of these stars. Perhaps the low value of $Y_0$ is controversial. While the adopted $Z_0$ of 16 Cyg B (0.024) is greater than the solar value, the $Y_0$ of all models (averaging 0.25) is much smaller than the solar value. This $Y_0$ is close to the primordial helium abundance \citep[0.2471;][]{2020A&A...641A...6P}.


The fact that the models have high $Z_0$ (0.020--0.025) and quite small $Y_0$ (0.24--0.26) leads us to doubt their validity. Taking a high $Z_0 $ increases the $M$ of the calibrated model. In this study, we set $Z_0 = 0.0177$. The model we obtained with $M$ = 0.97 $\MS$ and $R$ = 1.10 $\RS$ is in good agreement with observational constraints. For this well-fitting model, $Y_0 $ is 0.2825. The $\Dnu$, $\nu_0$, and $\delta \nu_{02}$ of this model are in excellent agreement with the observational values. The difference between the observational $\nu_0$ and the model $\nu_0$ (13 $\mu$Hz) is much smaller than $\Dnu$ (116.9 $\mu$Hz), while the model and observational $\delta \nu_{02}$ are close.



\subsubsection{KIC 6278762}
KIC 6278762, with an estimated age of 11.98 Gyr, is the oldest star in our sample and stands out as one of the oldest known planet-host systems. The observational asteroseismic parameters are high-quality, and the spectroscopic [M/H] and $T_{\rm eff}$ values reported in the literature show strong consistency. For this star, the observed $T_{\rm eff}$ is $5046 \pm 74$ K, in excellent agreement with our model value (5046 K), further supporting the reliability of the adopted model.

\subsubsection{70 Vir (TIC 95473936)}
Numerous spectral analyses have been conducted for this star. [M/H], $\log(g)$, and $T_{\rm eff}$ were determined in the analysis from 52 references in the literature. According to these data, there are linear relationships between $T_{\rm eff}$ and $\log(g)$, and between [M/H] and $\log(g)$. $\log(g)$ = 3.90, determined by asteroseismic data. The most suitable data for this value are [M/H] = -0.09 and $T_{\rm eff}$ = 5501 K \citep{2016A&A...587A.131M}.

We obtained five different solutions that are close to each other. The properties of these models are listed in Table \ref{tab:tic95473936}. Although the asteroseismic data are similar, only the input parameters $M$ and $Y_0$ differ. Model masses range from 1.001 to 1.059 $\MS$, and $Y_0$ ranges from 0.25907 to 0.28556. $Y_0$ decreases as the model mass increases. These models have two notable features. First, the model $\numax$ values are smaller than the observational $\numax$ values. Second, the age remains almost constant across different values of $M$ and $Y_0$. This demonstrates that such an analysis is successful in determining the age but limited in determining the chemical composition of such stars.

The star is close to $L_{\rm max}$ in the SG. It appears 200 K colder than that point.
The TIC catalogue gives $L = 3.038 \pm 0.081 \, {\rm L_{\odot}}$, and the model with the best fit for this luminosity has a mass of 1.059 ${\rm M_{\odot}}$. Table \ref{tab:tic95473936} demonstrates the significant strengths of asteroseismic modelling. While we can determine the age quite precisely, a good estimate of the chemical composition also requires a precise measurement of the observed luminosity. For this example star, $Z_{\rm 0}$ = 0.0117 and $Y_0$ = 0.25907. 
\begin{table}
    \centering
    \caption{Basic properties of four interior models for TIC 95473936. $Z_0$ and $\alpha$ of the models are 0.0117 and 1.8311, respectively. }
    \begin{tabular}{lllccccc}
        \hline
 $M$   &   $R$   & $Y_0$ & $T_{\rm eff}$ & $L$   & $\Dnu $ & $\numax$ & $t_9$ \\
 $\MS$ &   $\RS$ &       & K             & $\LS$   & $\mu$Hz & $\mu$Hz &  Gyr  \\
        \hline
1.001  &  1.869   & 0.28556  & 5493  & 2.865  &  53.80   &   912  & 8.44 \\
1.012  &  1.879   & 0.28056  & 5494   & 2.897  & 53.67   &   910  & 8.39 \\
1.034  &  1.897   & 0.27056  & 5498   & 2.963  & 53.61   &   908  & 8.29 \\
1.045  &  1.907   & 0.26556  & 5499   & 2.995  & 53.48   &   906  & 8.25 \\
1.059  &  1.919   & 0.25907  & 5499   & 3.037  & 53.33   &   904  & 8.19 \\
Obs.  &  -----   & -----  & 5501   & 3.038  &  53.67   &   939  & -----\\

        \hline
    \end{tabular}
    \label{tab:tic95473936}
\end{table}

%


\subsubsection{KIC 4143755}
This host is one of the lowest-mass stars in our catalogue. Literature data for this system are generally consistent. However, there is a significant difference between the model and spectral $T_{\rm eff}$ values. \cite{2016MNRAS.456.2183D} gives $T_{\rm eff}$ = 5622 and [M/H] = -0.4. The majority of the literature data agree with these values: $T_{\rm eff}$ = 5600 K and [M/H] = -0.5 dex. Obtaining a suitable fit with these values is difficult, and since the age is greater than the age of the Galaxy, our grid is insufficient. Our data is $T_{\rm eff}$ = 5746 K and [M/H] = -0.35 dex \citep{2016ApJ...822...86M}. With these values and using $\Dnu$=77.2 $\mu$Hz data, we obtained a model that agrees well with the observations by applying Method DZT: ($M$, $R$, $Z_{0}$, $t_{9}$, $T_{\rm eff}$)=(0.9078, 1.4210, 0.00819, 11.127, 5747). $\alpha$ and $Y_0$ of this model are 1.5652 and 0.26348, respectively. The observational reference frequencies of this host star are quite uncertain. There are only 5 points on the $\Dnu$-$\nu$ plot. In the models, $\nu_{\rm min0} \sim 1722$ and $\nu_{\rm min1}$=1126 $\mu$Hz.

\begin{figure}	
\includegraphics[width=1.15\linewidth]{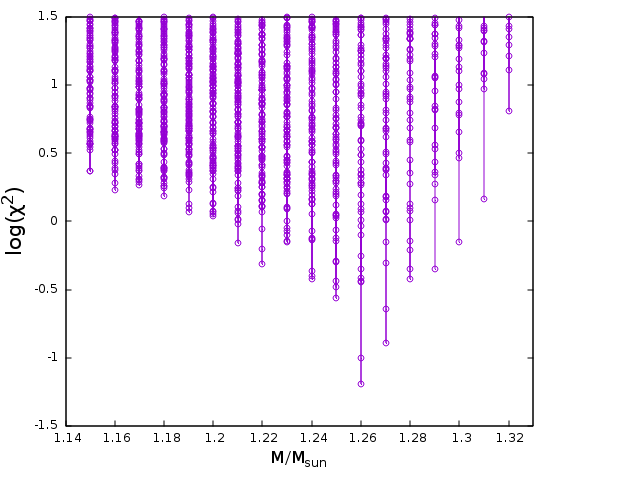}
\includegraphics[width=1.15\linewidth]{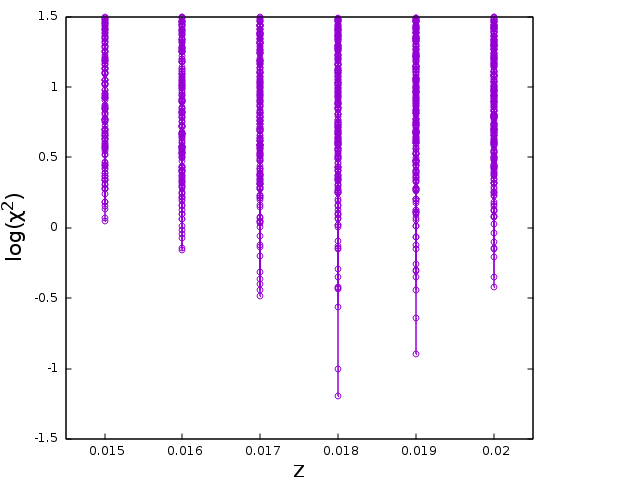}
\includegraphics[width=1.15\linewidth]{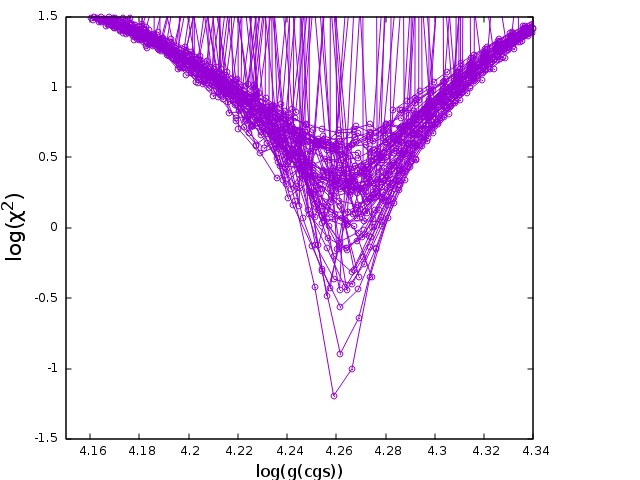}
    \caption{ $\log(\chi^2)$ is plotted with respect to $M$ (upper panel), $Z$ (middle panel), and $\log(g)$ (lower panel) for KIC 8866102. For each of the parameters, $\log(\chi^2)$ is minimized at a single point.}
    \label{fig:chi2_figure}
\end{figure}
\subsubsection{KIC 8866102}
Spectroscopic data for this host show significant scatter.
In this case, we can apply the MinZ method to KIC 8866102.
We calculated the $\Dnu$ of this star directly from the frequencies as 84.2 $\mu$Hz. It is given as 83.6 $\mu$Hz in the literature. 
For the MinZ method, $\chi^2$ is computed as in \cite{2025MNRAS.538..844O}. In Fig.\ref{fig:chi2_figure}(a), (b), and (c), $\log (\chi^2)$ is plotted with respect to
$M$, $Z$, and $\log g$, respectively. In all of the panels, $\chi^2$ is minimum at a single value of the horizontal axis. 
According to the MinZ method, ($M$, $Z$, $\log g$) = (1.25, 0.0163, 4.263). 
For a solar-like oscillating star with precise observational asteroseismic data, we obtain a unique model for the host star.  

This method is the most effective and consistent among those based on asteroseismic data, provided that the $\numin$ values are determined. The disadvantage of using $\numax$ as a constraint is that it can only be calculated from scaling relationships. $\Dnu$, on the other hand, is more useful because it can be calculated from both model and observational oscillation frequencies. However, in some cases, the average value of $\Dnu$ may depend on the frequency range used. $\numin$ values, however, can be calculated without resorting to scale relationships, and since there is no average value, there is no problem with choosing the frequency range used. If $\numin$ values can be obtained observationally, a unique model that best represents the star can be obtained using this method. The disadvantage of the method is that it requires the construction of numerous models. In this respect, the method needs further development.

\begin{figure}	\includegraphics[width=1.15\linewidth]{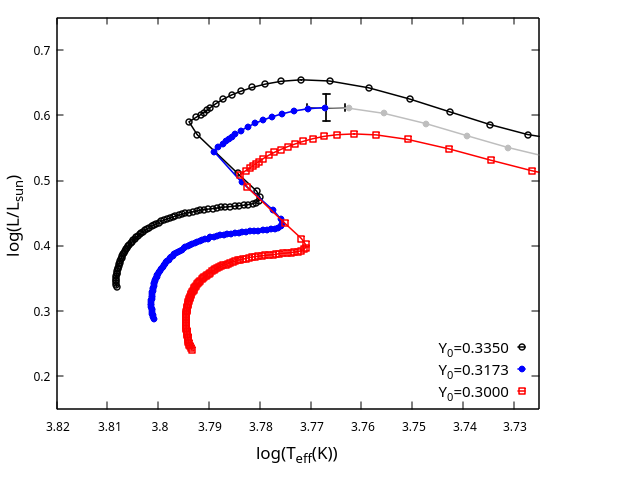}
    \caption{HRD of TIC 277890728 for the three models with different $Y_0$. The circle, filled circle, and box show the models with $Y_0$=0.3350, 0.3173, and 0.3000, respectively. The best model is with $Y_0$=0.3173}
    \label{fig:HRD_tic277890728_yVAR_figure}
\end{figure}
\subsubsection{TIC 277890728}
This star has a high Li abundance, A(Li)=2.65 \citep{2014A&A...562A..92D}. It is located in the region of the HR diagram where $L$ is at its maximum after TAMS. In this case, $\alpha$ becomes irrelevant in the $T_{\rm eff}$ calibration. By changing $Y_0$, we calibrate both $L$ and $T_{\rm eff}$. In Fig. \ref{fig:HRD_tic277890728_yVAR_figure}, the evolution traces of three different models constructed with three different $Y_0$s are plotted in HRD. The model parameters that best match the observed properties of the star are as follows: $M = 1.26$ $\MS$, $R = 1.97$ $\RS$, {$Y_{0} = 0.3173$, $Z_{0} = 0.0319$, and $t_{9}=4.75$ Gyr} (see Table \ref{table:model}).
%
%

\subsubsection{TIC 284181945}
The model parameters we obtained by applying the DZTLTeC method are: ($M$, $R$, $Z_{0}$, $t_{9}$, $T_{\rm eff}$)=(2.19, 10.33, 0.016, 0.76, 4894).
The $\Delta \Pi_1$ value for this star was found to be $303.96 \pm 0.30$ s by \cite{2024ApJ...971L..50L}. According to $\Dnu$ and $\Delta \Pi_1$, TIC 284181945 is an RC star. 

%

\subsubsection{TIC 38828538 (HD 29399)}
This host is an RGB star \citep{2024ApJ...971L..50L}.
$L$ is given by \cite{2022A&A...657A..89P} as 10.04 $\LS$.
3 separate solutions: (1) $L_{\rm mod}$=10.04, (2) $L_{\rm mod}$=11.6681, (3) $T_{\rm eff}$=4845 K ($L$=12.626 $\LS$).
$\Delta \Pi_1=83.55\pm0.09$ s \citep{2024ApJ...971L..50L}. The host is an RGB star.
\begin{table}
    \centering
    \caption{Basic properties of the models constructed for TIC 38828538.}
    \begin{tabular}{ccccc}
        \hline 
        $M$ ($\rm M_{\odot}$) & $R$ ($\rm R_{\odot}$) & $Z_{0}$ & $t_9$ (Gyr) & $T_{\rm eff}$ (K) \\
        \hline
        1.1631 & 4.6914 & 0.01856 & 6.7000 & 4738 \\ 
        1.2800 & 4.7100 & 0.01860 & 4.6600 & 4787 \\
        1.3663 & 4.9202 & 0.01868 & 3.7306 & 4806 \\ 
        1.4907 & 5.0431 & 0.01861 & 2.8387 & 4845 \\ 
        1.1394 & 4.4591 &   ---   &   ---  & --- \\ 
        \hline
    \end{tabular}
    \label{tab:TIC38828538}
\end{table}

When the difference between $T_{\rm mod}$ and $T_{\rm eff}$ was large, we obtained solutions with different $L$ values to test the uncertainty level in the luminosity. These solutions are listed in Table \ref{tab:TIC38828538}. $T_{\rm mod}$ increases with increasing L. When $L_{\rm mod}$=12.6 $\LS$, $T_{\rm mod}$ and $\teff$ are equal. In this case, because the mass increases significantly, there are substantial age differences. The last row of the table shows $M$ and $R$ calculated from corrected scaling relations. $\Dnu$ of the model with $M=$ 1.28 $\MS$ is in good agreement with the observed value. The difference between $T_{\rm mod}$ and $T_{\rm eff}$ is about 58 K for this model.



\subsubsection{TIC 233008631}
TIC 233008631 has the smallest values for $\Dnu$ and $\numax$ among the hosts. For such stars, asteroseismic modelling is complex because the frequencies of only a few modes can be calculated using the ADIPLS package. Therefore, $\Dnu$ can only be calculated over a narrow range. 


\subsubsection{Habitable planet around KIC 10593626}

This system, located at a distance of 197.5 pc, hosts a super-Earth at an orbital distance of 0.812 au \citep{2023A&A...677A..33B}. This planet, in the habitable zone, has an orbital period of 290 d. According to the model we obtained using the DZT method, the host's $Z_0$ is 0.0089. The age of the system was found to be 11.25 Gyr. If life had arisen, there would have been plenty of time for it to evolve. 

\section{Conclusions}
Asteroseismology is essential for understanding the internal structure of stars and constraining their fundamental parameters. Using quantities such as $\Dnu$, $\numax$, and $\delta \nu_{02}$, stellar properties can be determined with much higher precision than with other methods. In recent years, the growing number of solar-like oscillating host stars has further strengthened the link between asteroseismology and exoplanet research, enabling the two fields to advance together.

In this study, we compiled data on 127 host stars (and six candidates) from the literature and constructed interior models of these stars using the {\small MESA} code. Rather than applying a single method, we developed multiple approaches based on the available data and their associated uncertainties. We used $\Dnu$ and $Z_{\rm s}$ as constraints in all the methods we implemented. Additionally, we included one or more of 
$\delta \nu_{02}$, $\nu_{\rm min0}$, $\nu_{\rm min1}$, 
$L$, $T_{\rm eff}$, or $R$ as constraints in the applied methods, depending on the sensitivity of the data and their compatibility with other data.

Two key results emerged from our models. First, with only a few exceptions, the majority of host stars have a $Z_0$ greater than 0.007 (Fig. \ref{fig:Zt0}). This suggests that the threshold condition for planet formation is $Z_{0} \approx 0.007$; if the metallicity of a host star is below this value, no planet formation is observed.  Second, the age-metallicity relationship is quite variable. Both old and young stars can be metal-poor. For the most metal-rich stars of their time, the age--metallicity relationship appears linear up to $t_9$ = 6 Gyr. When $t_9$< 6 Gyr, it seems to remain constant in the range $Z_{0}$ = 0.007--0.033. These results provide important insights into the chemical evolution of the Galactic disc. To further examine this threshold, we modelled two hosts, TOI-2018 and HD 155358, reported to have very low metallicities, without asteroseismic constraints. Their surface metallicities ($Z_{\rm s} = 0.006$ and $0.004$) are below the critical value of $Z_0 \approx 0.007$. By including microscopic diffusion, we show that both stars can be reproduced with $Z_0 = 0.007$. The difference between $Z_0$ and $Z_{\rm s}$ is smaller in TOI-2018 and larger in HD 155358, reflecting the efficiency of diffusion. These results indicate that low surface metallicities do not necessarily imply low initial metallicities, and are consistent with the proposed threshold.

We also modelled 12 hosts classified as RC stars in the literature, taking into account mass-loss along the red giant branch. We show that mass loss has a crucial effect on the initial masses and ages of these stars, and may have important consequences for the survival of close-in planets during the RC phase.

For the solar-like oscillating KIC 8866102 with precise observational asteroseismic data ($\Dnu$, $\nu_{\rm min0}$, $\nu_{\rm min1}$) and $Z_{\rm s}$, we apply the MinZ method (Fig. \ref{fig:chi2_figure}) developed by \cite{2025MNRAS.538..844O}. We obtain a unique model for the host star. We could not apply this method to more stars due to the large number of models required.

Another important outcome of this study is the discovery of a relationship between $Z_0$ and the observed $Z_{\rm s}$ for the hosts. For the input parameter $Z_0$ used in the models, we derive a useful expression as a function of stellar M, R, and $Z_{\rm s}$. This expression can be used to estimate $Z_0$ by accounting for the decrease in $Z_{\rm s}$ caused by microscopic diffusion.

The radius of a planet is mainly dependent on its mass. It is known that the incident flux, in particular, has a significant effect on gas giants. When using irradiation energy per gram per second rather than flux, we obtain a one-to-one relationship for the radius. We show for the first time that $R_{\rm p}$ depends not only on $M_{\rm p}$ and $l_-$, but also on one of the orbital parameters ($P$ or $a$). Interestingly, our expression provides radius predictions for all planets, whether they are hot-Jupiter or terrestrial planets, including non-transiting planets.

This study provides the fundamental parameters of 127 solar-like oscillating host stars and their planets, establishing an essential catalogue in the literature. The TESS mission has significantly expanded the sample of evolved stars in the literature, allowing us to study a larger and more diverse sample of solar-like oscillating host stars. These systems serve as natural laboratories for understanding star–planet evolution and for investigating planet formation and disc evolution. The results offer valuable insights into key questions in both asteroseismology and exoplanet research, while laying a solid foundation for future studies.

Extended data for planetary systems are provided in NASA Exoplanet Archive \citep{2013PASP..125..989A} and TEPCat \citep{{2011MNRAS.417.2166S}}. It may be useful to create a continuously updated catalog of planetary systems with solar-like oscillating hosts using machine learning or to add asteroseismic data to existing catalogues.

\section*{Acknowledgements}
This work is supported by the Scientific and Technological Research Council of Turkey (T\"UB\.ITAK: 122F107, 123F019).

\section*{Data Availability}

 The data underlying this article will be shared on reasonable request to the corresponding author.



\appendix
\newpage
\onecolumn
\section{Basic asteroseismic and non-asteroseismic properties of {\it K\lowercase{epler}} and {\it TESS} Targets}

\small\addtolength{\tabcolsep}{-4pt}

\begin{landscape}

\begin{longtable}{@{}llccccccccccccccc@{}}
\caption[Continued.]{Observational physical properties of solar-like oscillating host stars and their planets. The columns are as follows, in order: host ID, effective temperature ($T_{\rm eff}$), host mass ($M$) and radius ($R$), luminosity ($L$), metallicity ([M/H]), large separation ($\Delta\nu$), frequency of maximum amplitude ($\nu_{\rm max}$), small separation ($\delta\nu_{02}$), reference frequencies ($\nu_0=\nu_{\rm min0}$ and $\nu_1 =\nu_{\rm min1}$), distance ($d$), semimajor axis ($a$), orbital period ($P$), planet radius ($R_{\rm p}$) and mass ($M_{\rm p}$). The final column lists the references. 
}\\
\hline
Host          &      $T_{\rm eff}$(K)     &      $M$($\MS$)           &       $R$ ($\RS$)           &      $L$ ($\LS$)           &   [M/H]     &       $\Delta \nu$ ($\mu$Hz)        &  $\nu_{\rm max}$ ($\mu$Hz)           &      $\delta \nu_{02}$  ($\mu$Hz)      &        $\nu_0$ ($\mu$Hz)      &       $\nu_1$ ($\mu$Hz)     &      $d$ (pc)    
   & $a$ (AU)     &      $P$ (day)     &       $R_{\rm p}$ (\RJ)   &     $M_{\rm p}$ (\MJ)    & Ref   \\
\hline
\endfirsthead

\caption[-- continued from previous page]{-- continued from previous page}\\
\hline
Host          &      $T_{\rm eff}$      &      $M$           &       $R$            &      $L$            &   [M/H]     &   $\Dnu$             &  $\numax$            &      $\delta \nu_{02}$        &        $\nu_0$       &       $\nu_1$      &      d    
   &a      &     $P$    &     $\Rp$    &    $\Mp$     & Ref   \\

\hline
\endhead
\hline
\endfoot
KIC   3544595 &  5669 $\pm$    75&   0.91$\pm$   0.03 &    0.92$\pm$   0.01 &    0.78$\pm$   0.02 &   -0.18$\pm$   0.01 &  145.77$\pm$   0.45 &  3366.0 $\pm$    81.0 &      8.6 $\pm$     1.7 &   3283.2 &  2701.9 &    96.0 &     0.053 &      4.727 &      0.1319 &     0.0126&  3,5,6,9,\textbf{79}  \\
KIC   3632418 &  6148 $\pm$   111&   1.41$\pm$   0.03 &    1.90$\pm$   0.02 &    4.87$\pm$   0.03 &   -0.19$\pm$   0.21 &   60.90$\pm$   0.40 &  1159.0 $\pm$    44.0 &      4.3 $\pm$     0.7 &   1422.1 &  1065.2 &   108.5 &     0.043 &      2.786 &      0.1462 &     0.024 & 6,8,9,\textbf{36},201  \\
KIC   4141376 &  6148 $\pm$    71&   1.03$\pm$   0.04 &    1.03$\pm$   0.02 &    1.33$\pm$   0.01 &   -0.24$\pm$   0.05 &  128.80$\pm$   1.30 &  2928.0 $\pm$    97.0 &          ---         &    ---   &   ---   &   214.4 &     0.103 &     11.873 &      0.1974 &     0.0160 &  5,9,\textbf{10}  \\
KIC   4143755 &  5746 $\pm$    66&   0.91$\pm$   0.06 &    1.41$\pm$  0.06&    1.93$\pm$   0.03 &   -0.35$\pm$   0.09 &   77.20$\pm$   1.30 &  1458.0 $\pm$    57.0 &          ---         &   1504.7 &  1272.7 &   385.7 &     0.135 &     19.56 &     0.220 &     --- &  5,9,\textbf{11},\textbf{37}  \\ 
KIC   4349452 &  6270 $\pm$    79&   1.17$\pm$   0.03 &    1.32$\pm$   0.02 &    2.46$\pm$   0.03 &   -0.04$\pm$   0.10 &   98.27$\pm$   0.57 &  2106.0 $\pm$    50.0 &      7.8 $\pm$     1.6 &   2365.2 &  1884.5 &   241.2 &     0.068 &      6.239 &      0.2452 &     0.0275 &  5,6,9,\textbf{82},201   \\
KIC   4914423 &  5845 $\pm$    88&   1.21$\pm$   0.03 &    1.49$\pm$   0.24 &    2.36$\pm$   0.05 &    0.07$\pm$   0.11 &   81.50$\pm$   1.60 &  1663.0 $\pm$    56.0 &          ---         &   1827.3 &  1581.1 &   506.3 &     0.133 &     15.965 &      0.311 &     0.0370 &   3,5,9,\textbf{83},202  \\
\end{longtable}
Ref. 1: \cite{2016MNRAS.456.2183D}, 2: \cite{2013ApJS..204...24B}, 3: \cite{2014ApJS..210...20M}, 4: \cite{2021A&A...649A...1G}, 5: \cite{2013ApJ...767..127H}, 6: \cite{2019MNRAS.490.1509K}, 7: \cite{2019ApJS..244...43Z}, 8: \cite{2014ApJS..210....1C}, 9: \cite{2020yCat.1350....0G}, 10: \cite{2018AJ....155..203H}, 11: \cite{2016ApJ...822...86M}, 12: \cite{2022ApJS..262...21F}, 13: \cite{2019AJ....157..245H}, 14: \cite{2019AJ....157..149L}, 15: \cite{2022ApJ...940...93D}, 16: \cite{2013ApJ...774...11K}, 17: \cite{2023PASJ...75.1030T}, 18: \cite{2013ApJ...771L..17B}, 19: \cite{2011ApJ...743..184W}, 20: \cite{2023AJ....166..167M}, 21: \cite{2021ApJS..255....8R}, 22: \cite{2024ApJ...971L..50L}, 23: \cite{2023A&A...678A.107P}, 24: \cite{2017AJ....153..136S}, 25: \cite{2005A&A...437L..31S}, 26: \cite{2021A&A...646A.131J}, 27: \cite{2025AJ....169...75S}, 28: \cite{2021MNRAS.502.3704A}, 29: \cite{2009ApJS..182...97W}, 30: \cite{2024AJ....168...27L}, 31: \cite{2019ApJ...885...31C}, 32: \cite{2014A&A...566A.113J}, 33: \cite{2016A&A...588A..62N}, 34: \cite{2018AJ....156..213M}, 35: \cite{2018MNRAS.479.1332M}, 36: \cite{2023A&A...677A..33B}, 37: \cite{2024ApJS..270....8W}, 38: \cite{2023A&A...669A..67H}, 39: \cite{2015A&A...573A..67P}, 40: \cite{2014A&A...561A..65B}, 41: \cite{2017A&A...599A..57B}, 42: \cite{2015A&A...575A..18B}  43: \cite{2011ApJS..197...26J}, 44: \cite{2013A&A...551A..90M}, 45: \cite{2012MNRAS.427..127B}, 46: \cite{2005A&A...440..609B}, 47: \cite{2022AJ....163..295B}, 48: \cite{2008PASJ...60.1317S}, 49: \cite{2016A&A...588A..98M}, 50: \cite{2015A&A...576A..94S}, 51: \cite{2020ApJ...896...65J}, 52: \cite{2017AJ....153...51W}, 53: \cite{2016AJ....152...19W}, 54: \cite{2017ApJS..230...12B}, 55: \cite{2015A&A...574A..50J}, 56: \cite{2021AJ....162...12V}, 57: \cite{2023RAA....23e5022X}, 58: \cite{2013PNAS..11013267G}, 60: \cite{2001A&A...375..205N}, 61: \cite{2018ApJ...868L..39H}, 62: \cite{2022AJ....163...79H}, 63: \cite{2022ApJS..259...45T}, 64: \cite{2015A&A...584A..79L}, 65: \cite{2012ApJ...744....4G}, 66: \cite{2024A&A...683A.125P}, 67: \cite{2020MNRAS.499.6084B}, 68: \cite{2016ApJS..225...32B}, 69: \cite{2006ApJ...646..505B}, 70: \cite{2009ApJ...703.1545F}, 71: \cite{2016ApJ...818...35W}, 72: \cite{2022A&A...663A...4S}, 73: \cite{2023Galax..11..112C}, 74: \cite{2010ApJ...721L.153J}, 75: \cite{2020A&A...641A..25N}, 76: \cite{2015A&A...573A...3J}, 77: \cite{2015ApJ...806...60K}, 78: \cite{2018AJ....156..264F}, 79: \cite{2015ApJ...800..135D}, 80: \cite{2024AJ....168..149B}, 81: \cite{2018A&A...620A..58S}, 82: \cite{2019AJ....157..145M}, 83: \cite{2019MNRAS.490.5103D}, 84: \cite{2014ApJS..210...25X}, 85: \cite{2015ApJ...799..170C}, 86: \cite{2016A&A...589A..93D}, 87: \cite{2014ApJS..210...19B}, 88: \cite{2015MNRAS.452.2127S}, 89: \cite{2022ApJ...926..120V}, 90: \cite{2023A&A...669A.117L}, 91: \cite{2013Natur.494..452B}, 92: \cite{2021MNRAS.507.1847R}, 93: \cite{2014ApJ...782...14V}, 94: \cite{2021AJ....162...89Z}, 95: \cite{2008ApJ...680.1450P}, 96: \cite{2013ApJ...766...40G}, 97: \cite{2012Sci...337..556C}, 98: \cite{2013ApJ...766..101C}, 99: \cite{2011ApJ...729...27B}, 100: \cite{2025arXiv250207996B}, 101: \cite{2022A&A...657A..87O}, 102: \cite{2019A&A...631A.136L}, 103: \cite{2008A&A...487..373S}  104: \cite{2015ApJ...806....5H}, 105: \cite{2013ApJ...762....9S}, 106: \cite{2018A&A...610A...3J}, 107: \cite{2015A&A...573A..36N}, 108: \cite{2012A&A...543A..54A}, 109: \cite{2010ApJ...713L.126B}, 110: \cite{2024ApJS..271...17Z}, 111: \cite{2004A&A...415.1153S}, 112: \cite{2023PASJ...75.1030T}, 113: \cite{2019A&A...622A.190A}, 114: \cite{2011AJ....141...16J}, 115: \cite{2017A&A...606A..51G}, 116: \cite{2019A&A...622A.190A}, 117: \cite{2022PASJ...74...92T}, 118: \cite{2008A&A...484L..21M}, 119: \cite{2018ApJ...860..109G}, 120: \cite{2018A&A...615A..31D}, 121: \cite{2014ApJ...785...94L}, 122: \cite{2022PASJ...74.1309T}, 123: \cite{2017AJ....153...21L}, 124: \cite{2016A&A...590A..38J}, 125: \cite{2011ApJ...736...87K}, 126: \cite{2021AJ....162..211H}, 127: \cite{2022AJ....164..156T}, 128: \cite{2022A&A...662A..12J}, 129: \cite{2022A&A...666A.125F}, 130: \cite{2016A&A...595A..55O}, 131: \cite{2020ApJ...898..119R}, 132: \cite{2011ApJ...727..117M}, 133: \cite{2018A&A...613A..47A}, 134: \cite{2024A&A...684A..85V}, 135: \cite{2009ApJ...701..154B}, 136: \cite{2025ApJ...978...24L}, 137: \cite{2018ApJ...861L...5G}, 138: \cite{2019RAA....19...41G}, 139: \cite{2018ApJ...866...99B}, 140: \cite{2013Sci...342..331H}, 141: \cite{2018A&A...616A...1G}, 142: \cite{2022A&A...657A..89P}, 143: \cite{2018A&A...616A..33S}, 144: \cite{2007A&A...475.1003H}, 145: \cite{2009ApJ...697.1263O}, 146: \cite{2021NatAs...5..775D}, 147: \cite{2016ApJS..224...12C}, 148: \cite{2018ApJS..239...32P}, 149: \cite{2015MNRAS.446.2959D}, 150: \cite{2017ApJ...835..172L}, 201: \cite{2022ApJS..259...35A}, 202: \cite{2016AJ....152....6W}, 203: \cite{2016A&A...594A..39F}, 204: \cite{2018MNRAS.481.3244G}, 205: \cite{2014ApJ...784...45R}, 206: \cite{2018ApJS..237...38B}, 209: \cite{2014A&A...570A..80T}, 210: \cite{2019ApJ...879...69T}, 211: \cite{2017AJ....154..107P}, 212: \cite{2022AJ....163..128W}, 213: \cite{2020A&A...633A..34C}, 214: \cite{2016A&A...587A.131M}, 215: \cite{2021A&A...647A.157S}, 216: \cite{2018A&A...614A..55A}, 217: \cite{2013A&A...556A.150S}, 220: \cite{2010MNRAS.403.1368G}, 221: \cite{2023ApJS..264...17Z}, 223: \cite{2023A&A...670A..73A}, 224: \cite{2019MNRAS.487.3162C}, 225: \cite{2022A&A...663A...4S}, 226: \cite{2023Natur.618..917H}, 227: \cite{2023ApJ...945...20J}, 228: \cite{2016ApJ...817...40F}, 229: \cite{2022PASJ...74...92T}, 230: \cite{2017AJ....154..274W}, 231: \cite{2007ApJ...657..546G}, 232: \cite{2018ApJS..235...38T}, 233: \cite{2014ApJ...789..154D}, 234: \cite{2010ApJ...720.1290G}, 235: \cite{2025MNRAS.541.2459W}, 236: \cite{2017MNRAS.469.1360C}, 237: \cite{2018ApJ...861..149F}

\end{landscape}

\small\addtolength{\tabcolsep}{-1pt}
\begin{longtable}{@{}llcccccccccccccccl@{}}
\caption[Continued.] {Basic physical properties of the interior models of the solar-like oscillating planet hosts and the revised mass and radius of the planets. {The reference frequencies are shown, defined as $\nu_0 = \nu_{\rm min0}$ and $\nu_1 = \nu_{\rm min1}$.}
}\\
\hline
       Host    & $T_{\rm eff}$&         $M$       &        $R$          &        $L$          &    $t_9$           &    $Z_0$ &   $Y_0$  &    $\alpha$  &   $Z_{s}$  &  $\Dnu$ &     $\numax$&$\delta \nu_{02}$&  $\nu_0$   &   $\nu_1$   & $R_{\rm p}$ &     $M_{\rm p}$ & Met. \\
              &  (K) &          $\MS$     &       $\RS$         &       $\LS$         & Gyr               &         &         &         &         &$\mu$Hz  &  $\mu$Hz &  $\mu$Hz &$\mu$Hz   & $\mu$Hz    &       $\RJ$   &    $\MJ$    & \\
    \hline
\endfirsthead

\caption[-- continued from previous page]{-- continued from previous page}\\
\hline
      Host    & $\teff$&         $M$       &        $R$          &        $L$          &    $t_9$           &    $Z_0$ &   $Y_0$  &    $\alpha$  &   $Z_{s}$  &  $\Dnu$ &   $\numax$&$\delta \nu_{02}$&  $\nu_0$   &   $\nu_1$   & $R_{\rm p}$ &     $M_{\rm p}$ & Met. \\ 
              &  (K) &          $\MS$     &       $\RS$         &       $\LS$         &  Gyr              &         &         &         &         &$\mu$Hz  &  $\mu$Hz &  $\mu$Hz &$\mu$Hz   & $\mu$Hz  &       $\RJ$   &    $\MJ$  & \\
\hline
\endhead
\hline
\endfoot
KIC   3544595 &  5624&    0.86$\pm$ 0.03 &     0.91$\pm$ 0.01 &    0.74 &    9.06$\pm$0.09 &  0.0111 &  0.2693 &  1.8213 &  0.0095$\pm$0.0002 &    145.8 &   3265.7 &      7.5 &   3273.7 &   2541.3 &   0.1302 &   0.0119 &  DZTref   \\
KIC   3632418 &  6088&    1.30$\pm$ 0.15 &     1.86$\pm$ 0.08 &    4.30 &    3.29$\pm$0.37 &  0.0163 &  0.2797 &  1.4599 &  0.0110$\pm$0.0041 &     61.0 &   1126.5 &      4.6 &   1402.4 &   1035.4 &   0.1435 &   0.0222 &  DZTref   \\
KIC   4141376 &  6151&    1.03$\pm$ 0.04 &     1.04$\pm$ 0.02 &    1.40 &    2.11$\pm$0.13 &  0.0088 &  0.2678 &  1.5980 &  0.0077$\pm$0.0008 &    128.6 &   2794.2 &     11.9 &   2992.0 &   2441.7 &   0.1996 &   0.0160 &  DZTL  \\
KIC   4143755 &  5747&    0.91$\pm$ 0.12 &     1.42$\pm$ 0.08 &    1.99 &   11.13$\pm$0.41 &  0.0082 &  0.2635 &  1.5652 &  0.0062$\pm$0.0011 &     77.2 &   1354.1 &     ---  &   1722.9 &   1126.1 &   0.2217 &     ---  &  DZTL  \\
KIC   4349452 &  6209&    1.10$\pm$ 0.05 &     1.28$\pm$ 0.02 &    2.20 &    5.39$\pm$0.14 &  0.0151 &  0.2779 &  2.2431 &  0.0130$\pm$0.0027 &     98.0 &   1961.4 &      6.1 &   2292.6 &   1715.4 &   0.2386 &   0.0259 &  DZTref   \\
KIC   4914423 &  5904&    1.17$\pm$ 0.07 &     1.49$\pm$ 0.04 &    2.42 &    4.99$\pm$0.18 &  0.0195 &  0.2861 &  1.7106 &  0.0168$\pm$0.0038 &     81.5 &   1604.4 &      5.3 &   1809.3 &   1391.0 &   0.3097 &   0.0356 &  DZTref   \\
KIC   5094751 &  5897&    1.02$\pm$ 0.08 &     1.32$\pm$ 0.05 &    1.89 &    7.94$\pm$0.25 &  0.0155 &  0.2781 &  1.8321 &  0.0127$\pm$0.0026 &     91.1 &   1770.6 &      5.7 &   1943.9 &   1492.6 &   0.2107 &   0.0140 &  DZT   \\
KIC   5866724 &  6211&    1.26$\pm$ 0.05 &     1.42$\pm$ 0.02 &    2.71 &    2.72$\pm$0.13 &  0.0237 &  0.2946 &  1.8311 &  0.0193$\pm$0.0033 &     89.9 &   1838.9 &      7.4 &   2084.9 &   1646.5 &   0.1273 &   0.0077 &  DZTref   \\
\end{longtable}
~~\\
\newpage
\twocolumn

\section{Planet hosts with very low metallicities}
\label{Appendix B}
\begin{figure}	
\includegraphics[width=1.15\linewidth]{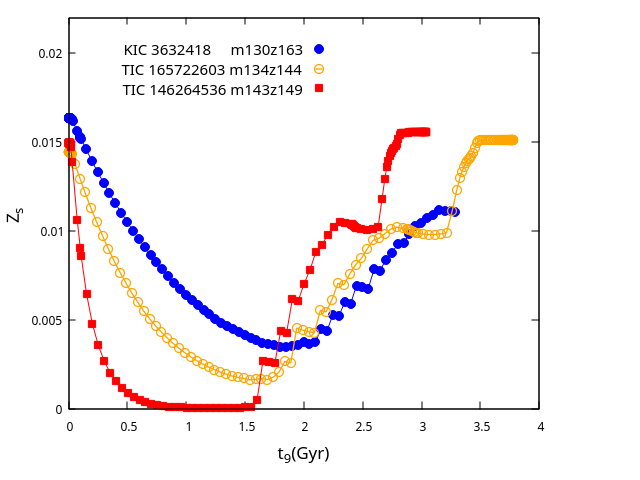}
\includegraphics[width=1.15\linewidth]{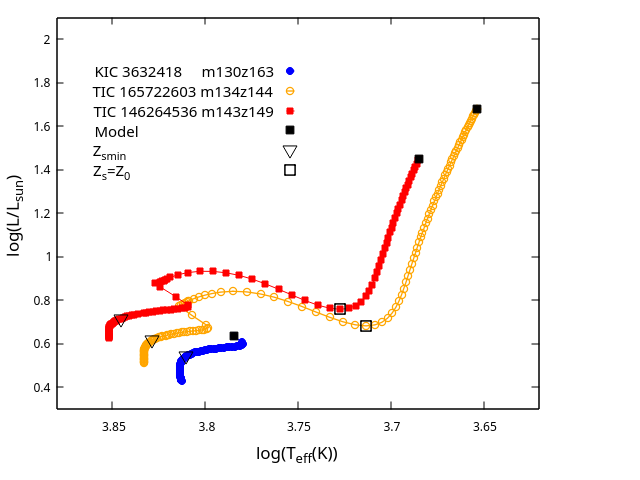}
    \caption{ a) Temporal evolution of $Z_{\rm s}$ for three planet hosts: KIC~3632418 (filled circles), TIC~165722603 (open circles), and TIC~146264536 (filled squares). For the latter two stars, the present-day $Z_{\rm s}$ remains close to $Z_0$ because these evolved objects experience the first dredge-up, which restores heavy elements previously depleted by diffusion. 
 b) Corresponding evolutionary tracks in the HR diagram. Filled symbols indicate the present-day positions of the stars, while triangles mark the evolutionary phase where $Z_{\rm s}$ reaches its minimum value. The box corresponds to the $Z_{\rm s}=Z_0$ case. 
    }
    \label{fig:Zs_t9_3hosts}
\end{figure}
\begin{figure}	
\includegraphics[width=1.15\linewidth]{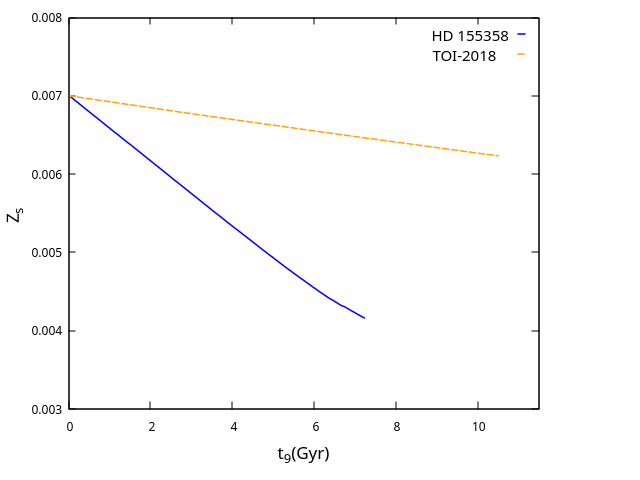}
    \caption{
    Evolution of the surface heavy-element abundance $Z_{\rm s}$ as a function of stellar age $t_9$ for TOI-2018 (dashed line) and HD~155358 (solid line). Both models adopt an initial heavy-element abundance $Z_0=0.007$. Due to microscopic diffusion, $Z_{\rm s}$ decreases to present-day values of 0.006 and 0.004, respectively.
    }
    \label{fig:Zs_t9_2hosts}
\end{figure}



Two planet hosts with very low metallicities reported in the literature are TOI-2018 and HD~155358, with [M/H] values of $-0.58$ dex \citep{2023AJ....166...49D} and $-0.60$ dex \citep{2017AJ....153..136S}, respectively. No asteroseismic constraints are currently available for these stars. Their observed [$\alpha$/Fe] values, 0.29 \citep{2023AJ....166...49D} and 0.10 \citep{2021MNRAS.505.4496G}, imply surface heavy-element abundances of $Z_{\rm s}=0.006$ and $Z_{\rm s}=0.004$, respectively. These values are lower than the critical initial metallicity obtained for the planet-host sample ($Z_0 \simeq 0.007$). To investigate whether these systems indeed formed with lower initial metallicities, stellar interior models including microscopic diffusion were constructed.

Among the 127 planet hosts analysed in this study, several stars exhibit present-day surface metallicities around $Z_{\rm s}\sim0.004$--0.005 despite having initial abundances $Z_0>0.007$. Three representative examples -- KIC~3632418, TIC~165722603, and TIC~146264536 -- experienced phases during their evolution in which $Z_{\rm s}$ became very low. Fig.~\ref{fig:Zs_t9_3hosts} shows the evolution of $Z_{\rm s}$ with age for these stars, all having $Z_0\simeq0.015$. The decrease in $Z_{\rm s}$ depends strongly on stellar mass. For KIC~3632418 ($1.30\,\MS$), $Z_{\rm s}$ decreases to $\sim0.004$ at $t_9\approx2$\,Gyr; for TIC~165722603 ($1.34\,\MS$), it reaches $\sim0.002$ at $t_9\approx1.5$\,Gyr; and for TIC~146264536 ($1.43\,\MS$), $Z_{\rm s}$ becomes nearly zero between $t_9=0.5$--1.5\,Gyr. These examples demonstrate that very low surface metallicities can naturally arise during stellar evolution even for relatively metal-rich initial compositions. This behaviour is also reflected in the evolutionary tracks shown in Fig.~\ref{fig:Zs_t9_3hosts}(b), where the minimum values of $Z_{\rm s}$ occur during the MS phase. As the stars evolve toward the RGB, the deepening convective envelope restores the surface composition, driving $Z_{\rm s}$ back toward its initial value $Z_0$.

The model properties of TOI-2018 and HD~155358 are listed in Table~\ref{table:mod_lowZ}, and the evolution of $Z_{\rm s}$ is shown in Fig.~\ref{fig:Zs_t9_2hosts}. Both stars are reproduced with an initial heavy-element abundance $Z_0=0.007$. The model age of TOI-2018 is 10.5 Gyr, while HD~155358 has an age of 7.23 Gyr. Owing to its low mass ($0.57\,\MS$), diffusion operates slowly in TOI-2018, resulting in a small difference between $Z_0$ and $Z_{\rm s}$ ($\sim0.001$). In contrast, diffusion is more efficient in HD~155358, producing a larger difference of $\sim0.003$. The evolutionary models of both stars are fully consistent with the results presented in Section \ref{sec: 4.6}. These results demonstrate that very low present-day surface metallicities do not necessarily imply low initial heavy-element abundances, and that stellar evolution effects, particularly microscopic diffusion, naturally reconcile the observed properties of these systems with a common initial metallicity threshold of $Z_0 \simeq 0.007$ for planet-host stars.

\twocolumn
\small\addtolength{\tabcolsep}{+2pt}

\begin{table}
    \centering

\caption{
Basic physical properties of the interior models of the planet hosts TOI-2018 and HD~155358, known for their very low surface metallicities. Surface abundances $Z_{\rm s}$ are derived from spectroscopic [M/H] and [$\alpha$/Fe] measurements and are found to be 0.006 and 0.004, respectively. For both stars, the models converge to an initial heavy-element abundance $Z_0=0.007$, consistent with the minimum metallicity inferred for the 127 solar-like oscillating planet hosts.
}
\begin{tabular}{llcccccccc}
\hline
       Host    & $T_{\rm eff}$&         $M$       &        $R$          &        $L$          &    $t_9$           &    $Z_0$ &   $Y_0$  &    $\alpha$  &   $Z_{s}$  \\
&  (K) &          $\MS$     &       $\RS$         &       $\LS$         & Gyr &         &         &         &         \\
TOI-2018 & 4306 &    0.57   &     0.57 &    0.10 &    10.50&  0.007 &  0.2611 &  1.8311 &  0.006   \\ 
Obs &  4302   &  --- &     --- &    0.10 &     --- &  ---  &  --- &  ---  &  0.006    \\
HD 155358 &  5954&    0.96&     1.33 &    2.01 &    7.23&  0.007 &  0.2611 &  1.3035 &  0.004    \\
Obs &  5958&    --- &     --- &    1.99 &     --- &  ---  &  --- &  ---  &  0.004    \\
\hline
\end{tabular}
\label{table:mod_lowZ}
\end{table}

\section{Comparison of revised mass and radius of planets and hosts - temporary section}
\label{app:C}
\begin{figure}	
\includegraphics[width=1.15\linewidth]{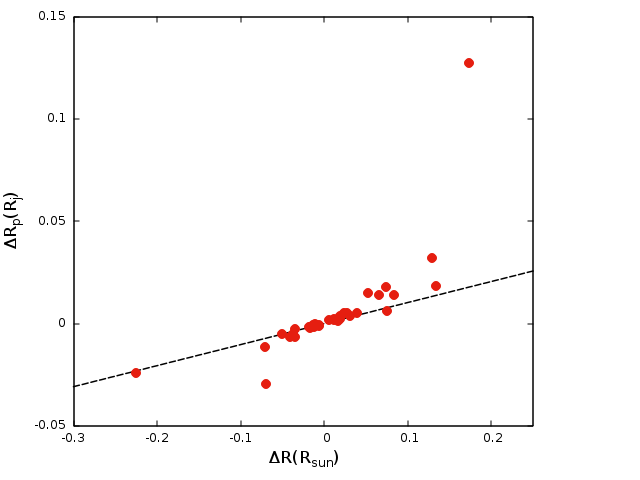}
\includegraphics[width=1.15\linewidth]{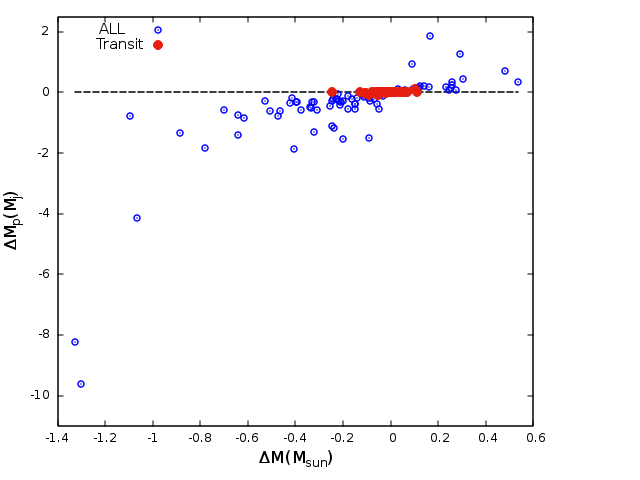}
    \caption{a) The difference between $R_{\rm p}$ and $R'_{\rm p}$ is plotted with respect to  $R$ and $R'$. b) The difference between $M_{\rm p}$ and $M'_{\rm p}$ is plotted with respect to  $M$ and $M'$.}
    \label{fig:dRMp_dRM}
\end{figure}

\bsp	
\label{lastpage}
\end{document}